\documentclass[12pt]{article}
\newcommand{\be}{\begin{equation}}
\newcommand{\ee}{\end{equation}}

\newcommand{\bi}{\begin{itemize}}
\newcommand{\ei}{\end{itemize}}
\newcommand{\bea}{\begin{eqnarray}}
\newcommand{\eea}{\end{eqnarray}}
\newcommand{\ba}{\begin{array}}
\newcommand{\ea}{\end{array}}

\usepackage[left=2.50cm, right=2.50cm, top=2.50cm, bottom=2.50cm]{geometry}
\usepackage[utf8]{inputenc}
\usepackage{cancel}
\usepackage[english]{babel}
\usepackage{physics}
\usepackage{hyperref}
\usepackage{amsthm}
\usepackage{mathrsfs}
\usepackage{mathtools}
\usepackage{graphicx}
\usepackage{changepage}
\usepackage{amssymb,amsmath}
\usepackage{verbatim}
\usepackage{empheq}
\usepackage{xcolor}
\usepackage{tikz}
\usepackage{float}
\usepackage{multirow}
\usepackage{multicol}
\usepackage{amsmath}
\usepackage{centernot}
\usepackage[hang,flushmargin]{footmisc} 
\usepackage[normalem]{ulem}
\usetikzlibrary{tikzmark}
\numberwithin{equation}{section}
\hypersetup{hidelinks}

\newlength{\bibitemsep}
\newlength{\bibparskip}
\let\oldthebibliography\thebibliography
\renewcommand\thebibliography[1]{%
  \oldthebibliography{#1}%
  \setlength{\parskip}{\bibitemsep}%
  \setlength{\itemsep}{\bibparskip}%
}
\begin{document}
\par
\bigskip
\Large
\noindent
{\bf 
Hall-like behaviour of higher rank Chern-Simons theory of fractons
\bigskip
\par
\rm
\normalsize

\hrule

\vspace{1cm}

\large
\noindent
{\bf Erica Bertolini$^{1,a}$},
{\bf Alberto Blasi$^{2,b}$}, 
{\bf Nicola Maggiore$^{2,3,c}$},\\
{\bf Daniel Sacco Shaikh$^{3,d}$}\\

\par

\small

\noindent$^1$ School of Theoretical Physics, Dublin Institute for Advanced Studies, 10 Burlington Road, Dublin 4, Ireland.

\noindent$^2$ Istituto Nazionale di Fisica Nucleare - Sezione di Genova, Via Dodecaneso 33, I-16146 Genova, Italy.

\noindent$^3$ Dipartimento di Fisica, Universit\`a di Genova, Via Dodecaneso 33, I-16146 Genova, Italy.
\smallskip

\smallskip

\vspace{1cm}

\noindent
{\tt Abstract}\\
Fracton phases of matter constitute an interesting point of contact between condensed matter and high-energy physics. The limited mobility property of fracton quasiparticles finds applications in many different contexts, including quantum information, spin liquids, elasticity, hydrodynamics, gravity and holography. In this paper we adopt a field theoretical approach to investigate the  three dimensional action of a rank-2 symmetric tensor field invariant under the covariant fracton symmetry. The theory appears as a non-topological higher rank generalization of the ordinary  Chern-Simons model, depending only on the traceless part of the tensor gauge field. After defining a field strength, a rank-2 traceless ``electric'' field and a ``magnetic'' vector field are identified, in analogy with the standard Chern-Simons ones. 
Once matter is introduced, a Hall-like behaviour with fractonic features emerges. In particular, our model shows a Hall-like dipole current, together with a vectorial ``flux-attachment'' relation for dipoles. This gives a possible starting point for a fracton - vortex duality. A gauge-fixing term is then introduced, from which propagators are computed and the counting of the degrees of freedom is performed. Finally, the energy-momentum tensor is shown to be conserved and the integrated energy density is proved to be zero, which reminds the topological nature of the standard Chern-Simons model.

\vspace{\fill}

\noindent{\tt Keywords:} \\
Quantum field theory, symmetric tensor gauge field theory, Chern-Simons theory, fractons.

\vspace{1cm}

\hrule
\noindent{\tt E-mail:
$^a$ebertolini@stp.dias.ie,
$^b$alberto.blasi@ge.infn.it,
$^c$nicola.maggiore@ge.infn.it,
$^d$4526348@studenti.unige.it.
}
\newpage

\section{Introduction}

Phases of matter have always inspired great interest in the scientific community. They describe emergent phenomena of the collective behaviour of a large number of strongly interacting particles, which are not obvious at the microscopic scale \cite{Anderson:1972pca}. To understand these complex physical phenomena a continuous interplay between Quantum Field Theory (QFT), condensed matter and statistical physics communities is necessary \cite{Sachdev:2012dq,Fradkin:2013sab,Witten:2015aoa}. Recently, the interest in a new kind of quasiparticle, known as ``fractons'' \cite{Vijay:2015mka,Vijay:2016phm,Nandkishore:2018sel,Pretko:2020cko}, has been growing more and more. Fractons are exotic excitations of certain quantum phases of matter which are completely immobile in isolation, and whose low energy physics can be described by higher rank theories. Together with fractons, which represent the extreme case, other quasiparticle excitations exist whose motion is limited to subdimensional spaces like lines or planes. These are sometimes called, respectively, ``lineons'' and ``planons''. Such constraints on motion typically derive from higher momentum conservation properties, the simplest of which is the dipole conservation, which have caught interest of the physics community both from an experimental and a theoretical point of view \cite{Nandkishore:2018sel,Pretko:2020cko,Caddeo:2022ibe,rasmussen,Pretko:2018jbi}. The first fracton models appeared as lattice theories in the context of quantum spin glasses \cite{Chamon:2004lew}, where the immobility property served as a tool to describe materials which cannot reach thermalization, and in quantum information \cite{Haah:2011drr,Bravyi:quantum} concerning the development of self-correcting quantum memories. There is not a unique theory for fractons and, in particular, depending on the quasiparticle content, these models were classified into two subclasses: ``type I'' and ``type II''. The former (type I) collects all the quasiparticle excitations (namely fractons, lineons and planons), while the latter (type II) only has fractons. Examples  are given, respectively, by the X-cube model \cite{Vijay:2015mka,Vijay:2016phm} and the Haah's code \cite{Haah:2011drr}. Since then, fractons have attracted more and more attention and quickly conquered many other areas of physics, including spin liquids \cite{Yoshida:2013sqa,Pretko:2016kxt}, elasticity  \cite{Pretko:2017kvd,Pretko:2019omh}, hydrodynamics \cite{Lucas,Ye,Wang,Surowska2,Surowska3},  holography \cite{Yan:2018nco}, gravity and  curved spacetimes \cite{Gu:2006vw,Xu:2006,Gu:2009jh,Xu:2010eg,Shirley:2017suz,Pretko:2017fbf,Jain:2021ibh,Tsaloukidis:2023bvz}. The origins of this surge in interest could be located in the development of gauge field theories for fractons \cite{Pretko:2016kxt,Pretko:2016lgv}. In particular, fractons are  described by higher rank (or higher spin) gauge theories, written in terms of a symmetric rank-2 field $h_{ij}(x)$ ($i,j$ spatial indices). These fractonic gauge theories closely reminds gravity theories \cite{Pretko:2017fbf} but show even stronger connections with the electromagnetic Maxwell theory, of which they represent an higher rank extension \cite{Pretko:2016lgv}. In these models, the limited mobility property is achieved through a Gauss-like constraint, which in the so-called ``scalar charge fracton theory'' is given by
\be \label{constraint}
    \partial_i\partial_j E^{ij}=\rho \ ,
\ee
where $E^{ij}(x)$ is a generalized rank-2 symmetric electric tensor field and $\rho(x)$ is the fractonic charge. 
In \cite{Pretko:2018jbi} the requirement \eqref{constraint} is encoded  within a generalized electromagnetism for a tensor gauge field $A_{ij}(x)$ transforming according to a higher rank generalization of the usual electromagnetic gauge transformation 
\begin{equation}
\delta A_{ij} = \partial_i\partial_j \phi\ .
\label{2}\end{equation}
The constraint \eqref{constraint} implies that the total dipole momentum
\be 
D^i\equiv \int dV\,x^i\rho\ 
\ee
must vanish when we integrate over an infinite volume
\be 
D^i=-\int dV\,\partial_jE^{ij}=0\ .
\ee
The fractonic charge $\rho(x)$ also satisfies the following generalized continuity equation
	\be\label{cont eq intro}
	\partial_{t}\rho+\partial_i\partial_jJ^{ij}=0\ ,
	\ee
where $J^{ij}(x)=J^{ji}(x)$ represents the dipole current \cite{Pretko:2016lgv}. This is another way of encoding dipole conservation 
\be\label{fracton dipole cons}
    \frac{dD^i}{dt}=\int dV x^i\frac{d\rho}{dt}=\int dV\partial_b J^{ib}=0\ ,
\ee
and thus another way of identifying fractonic behaviours. These conservation laws restrain completely the motion of single charges  (fractons), but keep dipoles free to move. Other models, also describing lineon and planon excitations, can be obtained from modifying the Gauss constraint (as in the ``vector charge fracton theory'') and/or adding constraints on the electric-like tensor field, such as tracelessness  for instance \cite{Pretko:2016kxt,Pretko:2016lgv}. The analogy with Maxwell theory also extends to the Hamiltonian density of the system, which is required to be $E^2(x)+B^2(x)$, where now $B^{ij}(x)$ is a magnetic-like rank-2 tensor, and to the rank-2 extension of the Faraday and Amp\`ere equations, together with the divergence of the magnetic-like field \cite{Pretko:2016lgv}. Notice that lattice and gauge fracton models  are physically different, describing, for instance, massive excitations in the first case and massless ones in the second, possibly related through a Higgs-like mechanism \cite{Ma:2018nhd,Bulmash:2018lid}. The continuum QFT limit of certain lattice models has also been studied \cite{Seiberg:2020wsg,Seiberg:2020bhn,Seiberg:2020cxy}, involving the concept of exotic global symmetries, which generate, for instance, the continuity equation \eqref{cont eq intro}. All the cases mentioned above are non-covariant descriptions of fracton models. From a pure quantum field theoretical point of view, in four spacetime dimensions (4D) a covariant theory of fractons has been shown \cite{Blasi:2022mbl,Bertolini:2022ijb,Bertolini:2023juh} to exist and to recover, within a coherent field theoretical set up, all the above \textit{ad hoc} constructions. Fracton properties can thus be derived from symmetry principles starting from the covariant extension of the rank-2 gauge field $h_{\mu\nu}(x)$ and looking for the most general 4D action invariant under the covariant fracton transformation 
	\be\label{fracton symm}
	\delta h_{\mu\nu}=\partial_\mu\partial_\nu\phi\ ,
	\ee
which generalizes \eqref{2}.
This single request leads to the above mentioned Maxwell-like equations, the fractonic Hamiltonian and a Lorentz-like force that acts on dipoles, but even more. For instance, the strong relation with gravity and, in particular, Linearized Gravity, which is receiving much attention in the community nowadays \cite{Afxonidis:2023pdq}, comes out naturally in that case, since the covariant fracton symmetry, also known as longitudinal diffeomorphisms \cite{Dalmazi:2020xou}, is a particular case of the diffeomorphism symmetry \cite{Hinterbichler:2011tt}. However, fracton models are not limited to 4D, but can also be found in other spacetime dimensions. In 3D, for instance, they can be found in the context of topological lattice defects. Indeed there exists an interesting duality which maps fractons into the theory of elasticity \cite{Pretko:2017kvd}. In this lower dimensional model the generalized magnetic field becomes a vector\footnote{We remind that the same happens in  Maxwell theory, where the magnetic field in 4D is a pseudovector, while in 3D is a pseudoscalar.} and this allows for a dictionary between the electric tensor field $E^{ij}(x)$ and the strain tensor $u^{ij}(x)$, and between the magnetic-like vector field $B^i(x)$ and the lattice momentum $\pi^i(x)$. Through this duality one can thus map defects of the material, such as disclinations and dislocations, into fractons and dipoles, respectively. A 3D fracton theory can also be recovered, from a QFT perspective, as the boundary of the 4D covariant one \cite{Bertolini:2023wie,Amoretti:2014iza}, where a Maxwell-Chern-Simons-like theory appears and a possible relation with higher order Topological Insulators exists through a generalized Kac-Moody algebra \cite{You:2019bvu,Bertolini:2023sqa}. A 3D boundary effect related to fractons was also studied in \cite{Pretko:2017xar} by adding a non-covariant, generalized theta-term to the 4D fractonic theories of \cite{Pretko:2016kxt,Pretko:2016lgv}. In that case a non-covariant Chern-Simons-like action was obtained and studied mostly in the context of a Witten effect for fractons and a possible generalization of Hall effect (also briefly analyzed in \cite{Prem:2017kxc}). Indeed, in 3D, higher rank Chern-Simons-like models are widely investigated. They can be found in the context of lattice theories and spin liquids \cite{Delfino:2022ndx,You:2019ciz,Delfino:2023anb}, and in relation to Hall systems \cite{Fliss:2021ekk,Cappelli:2015ocj}. For instance in \cite{Cappelli:2015ocj} a non covariant theory was studied with the aim of observing dipolar effects as higher order behaviours of the standard Hall theory description of Chern-Simons \cite{Dunne:1998qy,Tong:2016kpv}. Remarkably, this has been done before the development of fractons. Moreover, covariant higher rank Chern-Simons models appeared even earlier, in higher derivative gravity \cite{Bergshoeff:2009tb}, 
and in  \cite{Wu:1988py}, where the covariant transformation \eqref{fracton symm} was mentioned as a final comment. Although 3D fracton models appear, again, in many different areas of physics (lattice models \cite{Seiberg:2020bhn}, gravity theories \cite{Gromov:2017vir}, topological phases of matter and elasticity \cite{Pretko:2017kvd,Pretko:2019omh}), a systematic study from first principles and its physical consequences is still lacking. In this paper we shall study a 3D theory for fractons starting from a pure field theoretical perspective. Thus, as done in the 4D case \cite{Blasi:2022mbl,Bertolini:2022ijb,Bertolini:2023juh}, the \textit{covariant} fracton symmetry \eqref{fracton symm} will be the starting point to build the most general invariant, power-counting compatible action. In 3D there are two possible choices for the mass dimension of the gauge field, as it happens with the standard $U(1)$ vector gauge theory. In that case one can set the dimension of the field to be 1/2 and recover the topologically massive Maxwell-Chern-Simons theory \cite{Deser:1981wh}, while by setting the dimension to 1 one has the topological Chern-Simons theory \cite{Dunne:1998qy}. In this paper we will consider this last situation referred to fractons, thus we will study the theory of a rank-2 symmetric tensor field with mass dimension 1 and transforming under the covariant fracton symmetry.\\

The paper is organized as follows. In Section \ref{sec-puro fratt} the model is built as the most general invariant action, which looks like a higher-rank Chern-Simons theory, but is not topological and displays a dependence only on the traceless part of the gauge field. In analogy with the standard Chern-Simons theory and its relations with the Hall effect, generalized electric and magnetic field are identified. Starting from these definitions, in Section \ref{sec-currents} a coupling to matter is introduced in the theory and a fractonic behaviour is observed through a continuity equation displaying dipole conservation. The equations of motion are then studied, which lead to identify the dipole current as a Hall-like current. Furthermore, an equation of ``dipole-flux attachment'' emerges, allowing for a possible starting point of a fracton (dipole)-vortex duality as a higher rank extension of the standard particle-vortex duality. In Section \ref{sec-gauge} the gauge-fixing procedure is performed, and the propagators are computed. In Section \ref{sec-dof} the counting of the degrees of freedom (DoF) is done, and in Section  \ref{sec-emt} the energy-momentum tensor and its conservation are studied. Section \ref{sec-conclusion} concludes the paper with some final remarks. Appendix \ref{appQuant} discuss the level quantisation in more details, while Appendix \ref{appProp} reports the explicit computation of the propagators.

\vspace{1cm}

{\bf Notations}

3D=2+1 spacetime dimensions .\\
Indexes:  $\mu,\nu,\rho,...=\{0,1,2\}\ \,\ i,j,k,...=\{1,2\}$ .\\
Minkowski metric: $ \eta_{\mu\nu}=\mbox{diag}(-1,1,1)\ . $ \\
Levi-Civita symbol: $\epsilon_{012}=1=-\epsilon^{012}\ $ and
	\be
	\epsilon^{\mu\nu\rho}\epsilon_{\alpha\beta\gamma}=-\left|\begin{array}{ccc}
	\delta^\mu_\alpha&\delta^\mu_\beta&\delta^\mu_\gamma\\
	\delta^\nu_\alpha&\delta^\nu_\beta&\delta^\nu_\gamma\\
	\delta^\rho_\alpha&\delta^\rho_\beta&\delta^\rho_\gamma
	\end{array}\right|
		\quad\Rightarrow\quad\epsilon_{\mu\nu\rho}\epsilon^{\alpha\beta\rho}=-(\delta^\alpha_\mu\delta^\beta_\nu-\delta^\alpha_\nu\delta^\beta_\mu)
\quad\Rightarrow\quad
\epsilon^{\alpha\beta\mu}\epsilon_{\alpha\beta\nu}=-2\delta^\mu_\nu\ .
	\ee

\section{The model\label{sec-puro fratt}}

\subsection{The action}

We consider the 3D theory of a symmetric rank-2 tensor field $h_{\mu\nu}(x)$ 
\be
h_{\mu\nu}=h_{\nu\mu}
\label{}\ee
with mass dimension one
\be
[h_{\mu\nu}] = 1
\label{dimh}\ee
transforming under longitudinal diffeomorphisms  \cite{Dalmazi:2020xou}
\be\label{dfract1}
\delta h_{\mu\nu}=\partial_\mu\partial_\nu\phi\ ,
\ee
with $\phi(x)$ a local scalar parameter. The symmetry \eqref{dfract1} characterizes the covariant theory of fractons \cite{Bertolini:2022ijb}, 
and is affected by a kind of original sin. In fact, in 3D, due to \eqref{dimh}, the local gauge parameter $\phi(x)$ must have mass dimension $-1$. In order to have a gauge parameter $\phi(x)$ with vanishing mass dimension, as desirable, the tensor gauge field should have dimension two, which happens in six spacetime dimensions (for theories whose quadratic action has two spacetime derivatives, like Maxwell theory or Linearized Gravity) or in five dimensions (for theories displaying one derivative only, $i.e.$ Chern-Simons like theories). In lower dimensions, the gauge parameter unavoidably has negative dimension. This leads to singularities which manifest themselves only when treating the theory as a {\it gauge field theory}, as indeed it is, forcing to adopt the Landau gauge, as we shall see.
The most general 3D action invariant under \eqref{dfract1} and respecting power counting is 
	\be\label{Sinv1}
S_{inv}=\int d^3x\,\epsilon^{\mu\nu\rho}h_{\mu}^{\;\lambda}\partial_\nu h_{\rho\lambda}\ ,
	\ee
which depends only on the traceless part $\bar h_{\mu\nu}(x)$ of the tensor field 
$h_{\mu\nu}(x)$\footnote{From now on, we shall denote traceless tensors with a bar.}
\be\label{Sinv1traceless}
	S_{inv}=\int d^3x\,\epsilon^{\mu\nu\rho}\bar h_{\mu}^{\;\lambda}\partial_\nu\bar h_{\rho\lambda}\ ,
	\ee
where
\be
\bar h_{\mu\nu}\equiv h_{\mu\nu}-\frac{1}{3}\eta_{\mu\nu} h
\label{htilde}\ee
and
\be
h\equiv \eta^{\mu\nu}h_{\mu\nu}\ .
\label{h}\ee
The action $S_{inv}$ \eqref{Sinv1} for the rank-2 symmetric tensor field $h_{\mu\nu}(x)$ reminds the 3D Chern-Simons action for the ordinary vector gauge field $A_\mu(x)$~:
\be
S_{CS}[A] = \int d^3x\,\epsilon^{\mu\nu\rho}A_\mu\partial_\nu A_\rho\ ,
\label{CS}\ee
but, differently from the Chern-Simons action \eqref{CS}, the action \eqref{Sinv1traceless} depends on the metric and, hence, is not topological. Nonetheless, the generalized Chern-Simons action \eqref{Sinv1} shares with the ordinary one \eqref{CS} some peculiar properties. Indeed, as shown in  Appendix \ref{appQuant} following a procedure similar to the one adopted in \cite{Prem:2017kxc,Huang:2023zhp}, the same approach of ordinary Chern-Simons theory \cite{Dunne:1998qy,Tong:2016kpv} can be followed to show that the generalized Chern-Simons action for fractons is not invariant under large gauge transformations, but it changes by
\begin{equation}
\delta S_{inv} = 8\pi^2k\ ,
\end{equation}
where $k$ is the coupling constant of the generalized Chern-Simons action (which we reabsorbed by a redefinition of the field $h_{\mu\nu}(x)$). In order for the partition function 
\begin{equation}
Z[h_{\mu\nu}]=e^{iS_{inv}[h_{\mu\nu}]}
\end{equation}
to be invariant, it must be
\begin{equation}
k=\frac{n}{4\pi}\ ,\ \ n \in \mathbb{Z}\ ,
\end{equation}
which is the quantized level for dipolar Chern-Simons theory. Moreover, like ordinary Chern-Simons theory, the action \eqref{Sinv1} can be written using a gauge-invariant bulk theta-like term. In \cite{Bertolini:2022ijb}, it is introduced a 4D term
\begin{equation}
S_\theta = \int d^4x\, \theta\epsilon_{\mu\nu\rho\sigma}\partial^\mu h^{\lambda\nu}\partial^\rho h^\sigma_\lambda
\label{theta}\end{equation}
which is invariant under \eqref{dfract1}. As shown in \cite{Bertolini:2023sqa}, if $\theta$ is constant, \eqref{theta} is a boundary term, being a total derivative of our 3D action \eqref{Sinv1}. Finally in \cite{Bertolini:2022ijb} it has also been shown that this $\theta$-term can be written in terms of generalized 4D electric and magnetic tensor fields 
\begin{equation}
S_\theta\propto\int d^4x\, E^{il}B_{il}\ ,
\end{equation}
as conjectured in \cite{Pretko:2017xar}.
The equations of motion (EoM) derived from the topological action \eqref{CS} are
\be
\epsilon^{\mu\rho\sigma}F_{\rho\sigma} = 0\ ,
\label{CSeom}\ee
where $F_{\rho\sigma}(x)$ is the usual electromagnetic field strength
\be
F_{\rho\sigma} = \partial_\rho A_\sigma - \partial_\sigma A_\rho\ .
\label{emF}\ee
From the action \eqref{Sinv1traceless} the following EoM for the traceless tensor field $\bar h_{\alpha\beta}(x)$ are derived
	\be
	\frac{\delta S_{inv}}{\delta \bar h_{\alpha\beta}}=\epsilon^{\alpha\mu\nu}\partial_\mu \bar h_\nu^{\;\beta}+\epsilon^{\beta\mu\nu}\partial_\mu \bar h_\nu^{\;\alpha}=0\ ,\label{eom a fract}
	\ee
which also can be written in terms of a higher rank field strength, as we shall see shortly. 

\subsection{Generalized electric and magnetic fields}

As we remarked in the Introduction, rank-2 symmetric tensor fields are related to Linearized Gravity and to the more recent theory of  fractons \cite{Nandkishore:2018sel,Pretko:2020cko}, which shows a close analogy with electromagnetism \cite{Pretko:2016lgv,Bertolini:2022ijb}. The standard way to define the fracton tensor electric field, as done for instance in \cite{Pretko:2016lgv,Pretko:2017xar}, is to identify it with the spatial components of the conjugate momentum of $h_{\mu\nu}(x)$~:
\be
E^{ab}\equiv p^{ab} = \frac{\delta S_{inv}}{\delta \partial_t h_{ab}}\ .
\label{olddefE}\ee
This is motivated by the fact that in Maxwell theory the electric field can be equivalently defined by means of the electromagnetic field strength \eqref{emF}
\be
E^a\equiv F^{a0}
\label{standardE}\ee
or through the conjugate momentum $p^a(x)$
\be
E^a\equiv p^a=\frac{\delta S_{Max}}{\delta \partial_t A_{a}}\ .
\label{standardEp}\ee
The fact that in Maxwell theory the conjugate momentum coincides with the components $(a,0)$ of the field strength $F_{\mu\nu}(x)$ emphasizes the crucial property of the electric field of being a physical, gauge invariant, observable. In \cite{Bertolini:2022ijb} the generalized electric field $E^{ab}(x)$ has been defined as \eqref{olddefE} simply because a field strength generalizing the standard one $F_{\mu\nu}(x)$ was not known, at that time. The problem is that in 3D the conjugate momentum $p^{ab}(x)$ cannot be identified as an electric field, since
\be
p^{ab}=\frac{1}{2}\left(\epsilon^{0ac}h_c^{\;b}+\epsilon^{0bc}h_c^{\;a}\right)\ ,
\label{E=p}\ee
is not invariant under the transformation \eqref{dfract1}
\be
\delta p^{ab}=
\frac{1}{2}(\epsilon^{0ac}\partial^b\partial_c\phi +
\epsilon^{0bc}\partial^a\partial_c\phi)
\neq0\ .
\label{}\ee
and, hence, cannot represent a physical quantity
\be
E^{ab}\neq p^{ab}\ .
\label{Eneqp}\ee
This problem can be solved thanks to the higher rank field strength which has been defined in \cite{Bertolini:2022ijb}, in terms of which a coherent covariant formulation of the fracton theory has been given~:
\be
F_{\mu\nu\rho}= \partial_{\mu} h_{\nu\rho} + \partial_{\nu} h_{\mu\rho} 
-2 \partial_{\rho} h_{\mu\nu}\ .
\label{defF}\ee
The tensor $F_{\mu\nu\rho}(x)$ has the following properties~:
\begin{enumerate}
\item invariance
\be
\delta F_{\mu\nu\rho} = 0
\label{invF}\ee
\item symmetry
\be
F_{\mu\nu\rho}=F_{\nu\mu\rho}
\label{simF}\ee
\item cyclicity	
\be
F_{\mu\nu\rho} + F_{\rho\mu\nu} + F_{\nu\rho\mu} = 0
\label{ciclicitaF}\ee

\item Bianchi-like identity
\begin{eqnarray}
\epsilon_{\alpha\mu\nu\rho}\partial^\mu F^{\beta\nu\rho} &=& 0\qquad  \mbox{(4D)} \label{Bianchi4D}\\
\epsilon_{\mu\nu\rho}\partial^\mu F^{\alpha\nu\rho} &=& 0\qquad  \mbox{(3D)} \ .\label{Bianchi3D} 
\end{eqnarray}
\end{enumerate}
It is convenient to define a traceless field strenght $\bar F_{\mu\nu\rho}(x)$, reflecting the fact that the action $S_{inv}$ \eqref{Sinv1traceless} does not depend on the trace $h(x)$ \eqref{h}
\be\label{defFtraceless}
	\bar F_{\mu\nu\rho}\equiv F_{\mu\nu\rho}-\frac{1}{4}\left(2\eta_{\mu\nu}F^\lambda_{\ \lambda\rho}-\eta_{\mu\rho}F^\lambda_{\ \lambda\nu}-\eta_{\nu\rho}F^\lambda_{\ \lambda\mu}\right)\ ,
	\ee
which is fully traceless, in the sense that 
\be\label{TrF=0}
\eta_{\mu\nu}\bar F^{\mu\nu\rho}=\eta_{\mu\rho}\bar F^{\mu\nu\rho}=\eta_{\nu\rho}\bar F^{\mu\nu\rho}=0\ .
\ee
Evidently, $\bar F_{\mu\nu\rho}(x)$ shares with the full tensor $F_{\mu\nu\rho}(x)$ the properties \eqref{invF}, \eqref{simF} and \eqref{ciclicitaF}.
In close analogy with the Chern-Simons theory, the EoM \eqref{eom a fract} can be written as
\be
\frac{\delta S_{inv}}{\delta \bar h_{\alpha\beta}} = 
\frac{1}{3} \left(
\epsilon^{\alpha\mu\nu} \bar F^\beta_{\ \mu\nu} + 
\epsilon^{\beta\mu\nu}\bar F^\alpha_{\ \mu\nu}
\right)=0\ ,
\label{eomF}\ee
whose time and space components are
\bi
\item $\alpha=\beta=0$
	\be
	\frac{\delta S_{inv}}{\delta \bar h_{00}}=\frac{2}{3}\epsilon^{0mn}\bar F^0_{\ mn}=0
\label{eomF00}	\ee
\item $\alpha=a,\ \beta=0$
\be
\frac{\delta S_{inv}}{\delta \bar h_{0a}} = \frac{\delta S_{inv}}{\delta \bar h_{a0}} = \frac{1}{3}
\left(
\epsilon^{0kl}\bar F^a_{\ kl} -\frac{3}{2} \epsilon^{am0}\bar F^0_{\ 0m}
\right) =0
\label{eomF0a}\ee
\item $\alpha=a,\ \beta=b$
\be
\frac{\delta S_{inv}}{\delta \bar h_{ab}} =\frac{1}{3}
\epsilon^{0ak}\left(\bar F^b_{\ k0}- \bar F^b_{\ 0k}
\right) +
\frac{1}{3}
\epsilon^{0bk}\left(\bar F^a_{\ k0}- \bar F^a_{\ 0k}
\right)=0\ ,
\label{eomFaboffshell}\ee
\ei
where in \eqref{eomF0a} the ciclicity property \eqref{ciclicitaF} has been used. 
Notice that \eqref{eomF00} implies that, on-shell, the traceless generalized field strength has the following additional symmetry
\be
\bar F_{0mn} = \bar F_{0nm}\ .
\label{simmF}\ee
Using \eqref{simmF} and the ciclycity property \eqref{ciclicitaF}, the EoM \eqref{eomFaboffshell} reads
\be
\left.\frac{\delta S_{inv}}{\delta \bar h_{ab}}\right |_\eqref{simmF} =\frac{1}{2} \left(\epsilon^{0ak}\bar F^b_{\ k0}+
\epsilon^{0bk}\bar F^a_{\ k0}\right)=0\ ,
\label{eomFab}\ee
which will be useful in what follows. 
As anticipated, the higher rank field strength $F_{\mu\nu\rho}(x)$ allows to define
invariant higher rank generalizations of the electric field \eqref{standardE} and of the magnetic field.
In order to do so, we proceed, again, in analogy to what happens in Chern-Simons theory, where \cite{Dunne:1998qy,Tong:2016kpv}
\begin{eqnarray}
\frac{\delta S_{CS}[A]}{\delta A_a} &\propto& \epsilon^{0ab}E_b \label{tongdefE} \\
\frac{\delta S_{CS}[A]}{\delta A_0} &\propto& B \label{tongdefB}\ ,
\end{eqnarray}
where $B(x)$ is the the magnetic field which, in 3D, is a scalar quantity~:
\be
B\propto\epsilon^{0ab}F_{ab}\ .
\label{}\ee
We are therefore led to identify the higher rank generalizations of the standard definitions \eqref{tongdefE} and \eqref{tongdefB} as
\begin{eqnarray}
\left.\frac{\delta S_{inv}}{\delta\bar h_{ab}}\right|_\eqref{simmF} 
&\equiv& 
\frac{1}{2}\left(\epsilon^{0ak}\bar E^b_{\ k} + \epsilon^{0bk}\bar E^a_{\ k}\right)
\label{defE'} \\
\frac{\delta S_{inv}}{\delta \bar h_{0a}}&\equiv&B^a\ .   \label{defB'}
\end{eqnarray}
Comparing with \eqref{eomF0a} and \eqref{eomFab} and using \eqref{simmF}, we get\footnote{It is easy to see that, as a consequence of the definition \eqref{defFtraceless}, the tensorial electric field is spatially traceless $\eta_{ab}\bar E^{ab}=0$.}
\begin{eqnarray}
\bar E_{ab} &=&\bar F_{ab0} 
\label{defE1} \\
B^a &=&\frac{2}{3}\epsilon^{0mn}\bar F^a_{\ mn}=\epsilon^{0ab}\bar F^c_{\ cb}\label{defB1}\ ,
\end{eqnarray}
which gives 
	\be
	\bar F^{abc}=\frac{1}{2}\left(\epsilon^{0ac}B^b+\epsilon^{0bc}B^a\right)\ .
\label{F=B}	\ee
While the traceless electric field $\bar E_{ab}(x)$ can be obtained just by comparing \eqref{eomFab} and \eqref{defE1}, the derivation of \eqref{defB1} requires some care. In particular, the tracelessness \eqref{TrF=0} and cyclicity \eqref{ciclicitaF} properties of 
$\bar F_{\mu\nu\rho}(x)$ must be used. 
Because of the invariance of $\bar F_{\mu\nu\rho}(x)$, both $\bar E^{ab}(x)$ and $B^a(x)$ have the required property of being invariant 
\be
\delta_{fract} \bar E^{ab} = \delta_{fract} B^{a} =0\ .
\label{fractinvEB}\ee
Moreover, the tensorial traceless electric field $\bar E^{ab}(x)$ \eqref{defE1} turns out to be symmetric
\be
\bar E^{ab}=\bar E^{ba}\ ,
\label{simE}\ee
as it should be in fractonic theories \cite{Nandkishore:2018sel,Pretko:2020cko,Pretko:2016kxt,Pretko:2016lgv,Bertolini:2022ijb}. Notice that the traceless tensorial electric field $\bar E^{ab}(x)$ is symmetric as a consequence of \eqref{defE1} and not by definition. 
 
\section{Currents and fractons}\label{sec-currents}

Following \cite{Dunne:1998qy,Tong:2016kpv}, matter current  is introduced in Chern-Simons theory  by adding a term in the action
	\be
	S_{tot}=S_{CS}-\int d^3x A_\mu J^\mu\ ,
\label{CSmatter}	\ee
so that
	\be\label{J=eomCS}
	J^\mu=\frac{\delta S_{CS}}{\delta A_\mu}\ ,
	\ee
which encodes the matter response to electric and magnetic fields, through \eqref{tongdefE} and \eqref{tongdefB}. For instance, the time component of \eqref{J=eomCS} relates the magnetic flux to the electric density charge $J^0(x)$. The higher rank generalization of \eqref{CSmatter} is
	\be\label{Sinvmatter}
	S_{tot}=S_{inv}-\int d^3x\bar J^{\mu\nu}\bar h_{\mu\nu}\ ,
	\ee
where $S_{inv}$  is given by \eqref{Sinv1traceless}, and \eqref{J=eomCS} translates into
	\be\label{J=eom}
	\bar J^{\alpha\beta}=\frac{\delta S_{inv}}{\delta \bar h_{\alpha\beta}}\ .
	\ee
Recalling that we are working partially on-shell, since both the electric \eqref{defE1} and the magnetic field \eqref{defB1} are defined on the solution \eqref{simmF} of the EoM of $\bar h_{00}(x)$  \eqref{eomF00}, we have that 
	\be\label{J00=0}
	\bar J^{00}=0\ .
	\ee
Taking this into account, and using \eqref{defE'} and \eqref{defB'}, equation \eqref{J=eom} implies the following identifications
\begin{align}
\bar J^{0a}&= B^a \label{J=B}\\
\bar J^{ab}&= \frac{1}{2}\left(\epsilon^{0ak}\bar E^b_{\ k} + \epsilon^{0bk}\bar E^a_{\ k}\right)\label{J=E} \ .
\end{align}
Moreover, by writing explicitly \eqref{J=eom} 
\be\label{J=eom'}
	\bar J^{\alpha\beta}=\epsilon^{\alpha\mu\nu}\partial_\mu\bar h^{\ \beta}_\nu+\epsilon^{\beta\mu\nu}\partial_\mu\bar h^{\ \alpha}_\nu\ ,
	\ee
one can see that 
	\be\label{ddJ=0}
	\partial_\alpha\partial_\beta\bar J^{\alpha\beta}=0\ ,
	\ee
which is the rank-two extension of what happens in Chern-Simons theory, where \eqref{J=eomCS} implies the continuity equation \cite{Dunne:1998qy,Tong:2016kpv,Gromov:2017vir}
	\be\label{dJ=0CS}
	J^\mu\propto\epsilon^{\mu\nu\rho}\partial_\nu A_\rho\quad\Rightarrow\quad \partial_\mu J^\mu=0\ .
	\ee
Indeed, given \eqref{J00=0} and defining
\be
\rho\equiv2\partial_a\bar J^{0a}\ ,
\label{fracton charge}\ee
the conservation law \eqref{ddJ=0} represents a continuity equation
	\be\label{cont-fratt}
	\partial_0\rho+\partial_a\partial_b\bar J^{ab}=0\ ,
	\ee
which is relevant in fractonic theories, since it implies dipole momentum
\be
D^i=\int dV\,x^i\rho=-2\int dV\,\bar J^{0i}
\label{dipole} \ee
conservation  \cite{Nandkishore:2018sel,Pretko:2020cko,Pretko:2016kxt,Gromov:2017vir}\\
\be
\frac{dD^i}{dt}=\int dV x^i\frac{d\rho}{dt}=-\int dVx^i\partial_a\partial_b\bar J^{ab}=\int dV\partial_b\bar J^{ib}=0\ ,
\ee
defining the fracton quasiparticle as an immobile object \cite{Nandkishore:2018sel,Pretko:2020cko,Pretko:2016kxt}. In particular, it is possible to identify $\rho(x)$ as the charge density, $\bar J^{0a}(x)$ as the local dipole momentum and $\bar J^{ab}(x)$ as the dipole current, in agreement with \cite{Bertolini:2022ijb,Gromov:2017vir}. Notice that in \eqref{J=E} the current $\bar J^{ab}(x)$ coincides with the generalized Hall response described in \cite{Pretko:2017xar,Prem:2017kxc,Gromov:2017vir}, while \eqref{J=B} relates the magnetic vector field to the dipole charge. It is also interesting to notice that from \eqref{J=B} the presence of a nonvanishing divergence of the magnetic field implies a nonzero charge density 
\be
\rho=2\partial_a B^a\ ,
\label{defrho}\ee
which could be associated to the presence of a 3D fracton vortex defect.
Indeed, as remarked in \cite{Gromov:2017vir} in the context of a Chern-Simons-like theory with torsion, Eqs.\eqref{J=eom'} and \eqref{ddJ=0} can be seen as the starting point for the generalization of the particle-vortex duality for fractons. In fact the Chern-Simons current \eqref{dJ=0CS} and the Chern-Simons term \eqref{CS} are the main ingredients of the standard duality \cite{Dunne:1998qy,Tong:2016kpv}. The standard abelian Chern-Simons theory acquires a physical content only when coupled to matter fields, in the sense that otherwise the EoM \eqref{CSeom} has pure gauge solutions $F_{\mu\nu}=0$ only, which imply that the electric and magnetic fields vanish \cite{Dunne:1998qy}. For instance Chern-Simons models coupled to symmetry-breaking scalar fields, like the abelian Chern-Simons-Higgs model, have vortex solutions \cite{Dunne:1998qy,Jackiw:1990aw}. Vorticity is related to the charge $Q$ of the model, and hence to the magnetic flux $\Phi$, through 
\be
Q=\int d^2xJ^0=\int d^2x B=\Phi
\label{Q-flusso}\ee 
as a consequence of the so called ``magnetic flux attachment'' 
	\be\label{flux attach}
	J^0=B
	\ee
coming from the 0-component of the Chern-Simons EoM \eqref{J=eomCS}. Indeed the effect of the Chern-Simons coupling is to attach the magnetic flux of $B(x)$ to the matter charge density $J^0(x)$ in such a way that the flux follows the charge density wherever it goes \cite{Dunne:1998qy}. A model with these characteristics has relations with anyons and shows a self-duality property where both the scalar field and the gauge field acquire masses which are equal, and shows an $N=2$ supersymmetry \cite{Dunne:1998qy}. Considering all the analogies between the standard Chern-Simons theory and the Chern-Simons-like one \eqref{Sinv1traceless}, it would be interesting to see if our fractonic case, when coupled to matter, shows an analogous duality. If it does, it should be related to dipoles, as remarked in \cite{Prem:2017kxc}. Indeed Eq. \eqref{J=B}, which involves the generalized magnetic field $B^a(x)$, is  the fractonic equivalent of the flux attachment relation \eqref{flux attach}. In particular, the equivalent of the charge is a vector, represented by the the dipole density
	\be\label{dipole def}
	d^i\equiv -2\bar J^{0i}\ ,
	\ee
being the total dipole momentum $D^i$ given by \eqref{dipole}, for which \eqref{J=B} can be written as
	\be\label{dipole attach}
	d^i=-2B^i\ .
	\ee
Thus in our case flux attachment is related to dipoles. This is not surprising, if we  think about the fundamental properties of fractons, for which the relation \eqref{dipole attach} could not involve the fractonic charge $\rho(x)$ \eqref{fracton charge} since, by definition, it is an immobile object. Additionally, the standard particle-vortex duality with Chern-Simons term can be related to the Hall effect \cite{Tong:2016kpv}. It is interesting to see that the Hall analogy can be extended also by noting that the current $\bar J^{ab}(x)$ \eqref{J=E} can be interpreted as a generalized Hall current, if we define a generalized conductivity $\bar\sigma^{abmn}$ as
	\be\label{HallJ}
	\bar J^{ab}=\bar\sigma^{abmn}\bar E_{mn} \ ,
	\ee
where
	\be\label{sigma tilde}
	\bar\sigma^{abmn}=\bar\sigma^{bamn}=\bar\sigma^{abnm}\equiv\frac 1 4 \left(\epsilon^{0am}\eta^{bn}+\epsilon^{0bm}\eta^{an}+\epsilon^{0an}\eta^{bm}+\epsilon^{0bn}\eta^{am}\right)\ ,
	\ee
which is the $mn$-symmetrization of the one found in \cite{Pretko:2017xar,Prem:2017kxc}. This mimics the standard Hall current $J^i=\sigma^{ij}E_j$ also in its orthogonality condition with the (generalized) electric field \eqref{defE1}
	\be\label{JE=0}
	\bar J^{ab}\bar E_{ab}=\epsilon^{0mn}\bar E_{ma}\bar E_n^{\ a}=0\ .
	\ee
We thus have from the EoM with matter coupling \eqref{J=eom} the ``dipole-flux attachment'' \eqref{dipole attach} and the generalized Hall current \eqref{HallJ}, which represent the two main ingredients for the fracton generalization mentioned above
	\begin{table}[H]
	\centering
		\begin{tabular}{c|cc}
		standard Chern-Simons&fractonic Chern-Simons&\\\hline
		$\rho=B$&$d^i=-2B^i$&flux attachment\\
		$J^i=\sigma^{ij}E_j$&$\bar J^{ab}=\bar\sigma^{abmn}\bar E_{mn}$&Hall(-like) current\\
		\end{tabular}
	\end{table}
Therefore this higher rank version, described by  the Chern-Simons-like action \eqref{Sinv1traceless}, could be the core of a generalized fracton-vortex duality, with possible consequences in the context of elasticity theory \cite{Gromov:2017vir} and of a fractonic Hall effect.
A comment is in order concerning the existing Literature. The invariant action \eqref{Sinv1} reduces to that studied in \cite{Prem:2017kxc} in the particular case $\bar{h}_{j0}=\partial_j \psi$, with $\psi(x)$ a scalar function. In the context of this paper, this is equivalent to work partially on-shell on the solution of the equation of motion of $\bar{h}_{00}$ \eqref{eomF00}, which plays a crucial role in the definition of both the electric \eqref{defE1} and the magnetic  \eqref{defB1}
fields. Moreover, it implies $\bar{J}^{00}=0$, hence leading us to the relevant fractonic continuity equation \eqref{cont-fratt}. Going back to the connections with the work \cite{Prem:2017kxc}, we observe that we recover their  generalized Hall response in symmetrized form. Remarkably, differently from \cite{Prem:2017kxc}, which starts directly by writing a  non-covariant Chern-Simons like action, we derive ours as the most general 3D action invariant under the covariant fractonic gauge transformation \eqref{dfract1}
 with $[h_{\mu\nu}(x)]=1$. Moreover, a rank-two Chern-Simons action appears also in \cite{You:2019ciz}, where the Authors start from a lattice model for a rank-2 symmetric off-diagonal spatial tensor gauge field $A_{ij}(x)$ with $i,j=1,2,3$. Hence, differently from the case treated in the present paper, they work in three spatial dimensions. In addition, \cite{You:2019ciz} is mainly dedicated to the study of generalized Chern-Simons theories with only two spatial gauge fields and quadratic in the derivatives appearing in the gauge transformations. This is different from what we have done, since it corresponds to a vector gauge theory.

\section{Gauge-fixing and propagators}\label{sec-gauge}

The most general gauge fixing for the transformation \eqref{dfract1}, which depends on the scalar local parameter $\phi(x)$, is 
	\be
\partial^\mu\partial^\nu h_{\mu\nu}+\kappa \partial^2 h = 0	\ ,
\label{gaugefixing}\ee
where $\partial^2=\partial^\mu\partial_\mu=\Box$ is the 3D D'Alembertian operator. The scalar constraint \eqref{gaugefixing}
is realized by adding to the action $S_{inv}$ \eqref{Sinv1} (or, equivalently, \eqref{Sinv1traceless}) the gauge fixing term
	\be\label{Sgf}
	S_{gf}=-\frac{1}{2\xi}\int d^3x\,
	\left(
	\partial^\mu\partial^\nu h_{\mu\nu}+\kappa \partial^2h
	\right)^2\ ,
	\ee
which depends on two gauge parameters $\xi$ and $\kappa$. The former, $\xi$, identifies the type of gauge fixing. For instance, $\xi=0$ and $\xi=1$ represent the Landau and Feynman gauges, respectively. 
The direct consequence of the fact that the local gauge parameter $\phi(x)$ in \eqref{dfract1} has negative dimensions, is that the constant gauge parameter $\xi$ in the gauge fixing term $S_{gf}$ \eqref{Sgf} is massive $[\xi]=3$, which leads to infrared divergences exactly like in ordinary Chern-Simons theory, which, for the same reason, is defined in the Landau gauge only \cite{Alvarez-Gaume:1989ldl,Blasi:1989mw,Delduc:1990je}
\be
\xi=0\ .
\label{Landaugauge}\ee 
In momentum space and in the Landau gauge $\xi=0$, the propagator of the gauge fixed action 
\be
S_{tot} = S_{inv} + S_{gf}\ ,
\label{Stot}\ee
 is
\be
\begin{split}
&\langle
\hat h_{\alpha\beta}(p)\hat h_{\rho\sigma}(-p)
\rangle
=\hat\Delta_{\alpha\beta,\rho\sigma}(p)=  \\
&\frac{ip^\lambda}{16p^2}\left[
\epsilon_{\alpha\lambda\rho}
\left(
\eta_{\beta\sigma} + 3 \frac{p_\beta p_\sigma}{p^2}
\right)
+
\epsilon_{\alpha\lambda\sigma}
\left(
\eta_{\beta\rho} + 3 \frac{p_\beta p_\rho}{p^2}
\right)
+
\epsilon_{\beta\lambda\rho}
\left(
\eta_{\alpha\sigma} + 3 \frac{p_\alpha p_\sigma}{p^2}
\right)
+
\epsilon_{\beta\lambda\sigma}
\left(
\eta_{\alpha\rho} + 3 \frac{p_\alpha p_\rho}{p^2}
\right)
\right] 
\end{split}
\label{proplandau}\ee
where $p^2\equiv p^\mu p_\mu$ and $\hat h_{\mu\nu}(p)$ is the Fourier transform of $h_{\mu\nu}(x)$ (the details of the calculation can be found in Appendix A). Notice that \eqref{proplandau} implies
\be
\langle \hat h(p)\hat h_{\rho\sigma}(-p) \rangle = 
\langle \hat h(p)\hat h(-p) \rangle = 0\ ,
\label{proptrace}\ee
as it should, being the theory traceless.
The propagator \eqref{proplandau}  does not display any pole and does not depend on the gauge parameter $\kappa$, coupled in the gauge fixing term $S_{gf}$ \eqref{Sgf} to the trace field $h(x)$, which is expected, being the theory defined by $S_{inv}$ \eqref{Sinv1traceless} traceless. The gauge fixing term \eqref{Sgf} can alternatively be written as
\be
S^{(\xi)}_{gf} = \int d^3x\,
\left[ b \left(
\partial^\mu\partial^\nu h_{\mu\nu}+\kappa \partial^2h\right) +\frac{\xi}{2}b^2
\right]
\label{Sgfb}\ee
where the scalar field $b(x)$ is the Nakanishi - Lautrup Lagrange multiplier implementing the gauge condition \eqref{gaugefixing}. This form of the gauge fixing term will be useful in the next Section, when we shall count the degrees of freedom.

\section{Degrees of freedom}\label{sec-dof}

To count the degrees of freedom (DoF) of the theory, we look for the number of independent components of the tensor field $h_{\mu\nu}(x)$ which, as a symmetric rank-2 tensor field, has a total of six components. The usual way to proceed \cite{Blasi:2015lrg,Blasi:2017pkk,Gambuti:2020onb,Gambuti:2021meo,Bertolini:2021iku,Maggiore:2017vjf} is to write the equations of motion of the gauge fixed action $S_{tot}$ \eqref{Stot} in momentum space, and it is convenient to write the gauge fixing term $S_{gf}$ in the form \eqref{Sgfb}, where the scalar gauge condition is implemented by the scalar Lagrange multiplier $b(x)$. As already said, the Landau gauge \eqref{Landaugauge} is mandatory, due to fact that the gauge parameter $\xi$ is massive $([\xi]=3)$. Therefore, the gauge fixing term is
\be
S_{gf}^{(\xi=0)} = 
\int d^3x\,
b \;
(\partial^\mu\partial^\nu h_{\mu\nu}+\kappa\partial^2h)
  \ .
\label{Sgfblandau2}\ee
In momentum space, the EoM of the theory are
\bea
\frac{\delta S_{tot}}{\delta \hat h_{\mu\nu}} &=&
i \epsilon^{\alpha\beta\mu} p_\beta\hat{\bar h}^\nu_{\ \alpha} +
i \epsilon^{\alpha\beta\nu} p_\beta\hat{\bar h}^\mu_{\ \alpha} - p^\mu p^\nu \hat b 
-\kappa\eta^{\mu\nu}p^2\hat b
\label{eomhp2traceless}\\
\frac{\delta S_{tot}}{\delta \hat b} &=&  -p^\mu p^\nu {\hat{h}}_{\mu\nu}-\kappa p^2\hat h= -p^\mu p^\nu {\hat{\bar h}}_{\mu\nu}-\left(\tfrac{1}{3}+\kappa\right) p^2\hat h
\label{eombp2}\ ,
\eea
where ${\hat{\bar h}}_{\mu\nu}(x)$ is the Fourier transform of the traceless part $\bar h_{\mu\nu}(x)$ of the tensor field $h_{\mu\nu}(x)$.
From the EoM \eqref{eomhp2traceless}, on shell one gets
\be
\left(\epsilon_{\mu\lambda\rho}\frac{p^\lambda p_\nu}{p^2}\right)
\frac{\delta S_{tot}}{\delta \hat h_{\mu\nu}} =
 -i\frac{p_\rho}{p^2}\left(p^\nu p^\lambda \hat{ h}_{\nu\lambda}\right) +
 ip^\nu\hat{h}_{\nu\rho}=
 -i\frac{p_\rho}{p^2}\left(p^\nu p^\lambda \hat{\bar h}_{\nu\lambda}\right) +
 ip^\nu\hat{\bar h}_{\nu\rho}=0\ .
\label{3cond2traceless}\ee
The theory defined by the covariant fracton symmetry \eqref{dfract1} is described by the invariant action $S_{inv}$ \eqref{Sinv1}, which actually  is traceless \eqref{Sinv1traceless}. The trace $h(x)$ is introduced only through the gauge fixing term $S_{gf}$ \eqref{Sgf}, which is needed to well define the Green function's generating functional $Z[J]$. The role of the gauge fixing procedure is that of eliminating the redundant degrees of freedom, and certainly not that of introducing new ones. Therefore, the trace $h(x)$, which enters the theory only through the gauge fixing term, cannot count as a physical DoF. The outcome of this reasoning is that, in order to avoid to introduce the trace $h(x)$ as a spurious DoF, we may (we should, actually), set in $S_{gf}$ \eqref{Sgf} the gauge parameter $\kappa$ to the value which corresponds to gauge fix only the traceless part $\bar h_{\mu\nu}(x)$ of $h_{\mu\nu}(x)$, that is  
	\be
	\kappa=-\frac{1}{3}\ ,
	\ee
so that
\be
\left. S_{gf}^{(\xi=0)}\right|_{\kappa=-\frac{1}{3}} =
\int d^3x\,
b \;
\partial^\mu\partial^\nu \bar h_{\mu\nu}
  \ .
\label{Sgftrace}\ee
Using the momentum space gauge condition
	\be
	\frac{\delta S_{tot}}{\delta \hat b} = -p^\mu p^\nu {\hat{\bar h}}_{\mu\nu}=0\ 
	\ee
in \eqref{3cond2traceless}, gives
\be\label{ph=0}
p^\nu\hat{\bar h}_{\nu\rho}=0 \,
\ee
representing three conditions on the five components of the traceless rank-2 symmetric tensor field $\hat{\bar h}_{\nu\alpha}(p)$, which therefore has two independent components. Hence, the degrees of freedom of the theory are two.

 \section{Energy momentum tensor}\label{sec-emt}
 
 Chern-Simons and BF theories are topological QFTs of  the Schwarz type  \cite{Birmingham:1991ty}, characterized by an invariant action which does not depend on the metric $g_{\mu\nu}(x)$. The energy momentum tensor of topological QFTs vanishes
 \be
 T_{\mu\nu}= -\frac{2}{\sqrt{-g}}\frac{\delta S_{inv}}{\delta g^{\mu\nu}}=0\ .
 \label{Tmunu}\ee
This renders the energy momentum tensor unphysical, in the sense that the only contribution to it comes from the gauge fixing term, which depends on the metric. One of the most striking consequences of this fact is that topological QFTs have vanishing energy density
\be
T_{00}=0\ ,
\label{T00=0}\ee
and hence vanishing energy. The model presented in this paper is a higher rank generalization of the standard Chern-Simons theory, and it is not topological. In fact, $S_{inv}$ \eqref{Sinv1} displays a metric dependence
\be
S_{inv} = \int d^3x\; \epsilon^{\mu\nu\rho}g^{\alpha\beta}
h_{\alpha\mu}\partial_\nu h_{\beta\rho}\ ,
\label{Sinvconmetrica}\ee
which is mild, being only linear. To make a comparison, Maxwell theory has a cubic dependence on the metric
\be
S_{Max}=-\frac{1}{4}\int d^4x\sqrt{-g}\;g^{\mu\nu}g^{\rho\sigma}F_{\mu\rho}F_{\nu\sigma}\ .
\label{maxconmetrica}\ee
We might say that the theory described by the action $S_{inv}$ \eqref{Sinvconmetrica} is ``almost'' topological. There is indeed at least one property which our theory shares with a topological QFT: it has, on-shell, vanishing energy. In fact, the energy momentum is
\be
T_{\mu\nu}=-\epsilon^{\alpha\beta\gamma}
(h_{\alpha\mu}\partial_\beta h_{\gamma\nu}+ h_{\alpha\nu}\partial_\beta h_{\gamma\mu})\ ,
\label{}\ee
whose 00-component, namely the energy density, is
\be\label{T00}
	T_{00}=-2\epsilon^{\mu\nu\rho}h_{\mu0}\partial_\nu h_{\rho0}=-2\epsilon^{0mn}\left(h_{00}\partial_mh_{n0}-h_{m0}\partial_0h_{n0}+h_{m0}\partial_nh_{00}\right)\ .
	\ee
The EoM \eqref{eomF00} of $\bar h_{00}$ is
\be
	\epsilon^{0mn}\partial_mh_{0n}=0\ ,
\ee
which is solved by
	\be\label{h0isol}
	h_{0i}=\partial_i\psi\ ,
	\ee
where $\psi(x)$ is a generic scalar field. It is immediate to see that on this solution (corresponding to \eqref{simmF}), the total energy, given by the volume integral of the energy density \eqref{T00}, vanishes
\be
\int dV\,T_{00}=0\ ,
\label{}\ee
which is a weaker statement of the corresponding one concerning topological QFTs.  One can also verify that the energy momentum tensor is conserved on-shell~:
\be
\partial^\alpha T_{\alpha\beta}=0\ .
\label{Tconsx}\ee
To prove this, it is convenient to work in momentum space, where \eqref{Tconsx} reads
\be
p^\alpha \hat T_{\alpha\beta}=0\ .
\label{Tconsp}\ee
In fact
	\be
		\begin{split}
		p^\alpha\hat T_{\alpha\beta}&=i\epsilon^{\mu\nu\rho}p_\nu\left[p^\alpha\hat{\bar h}_{\mu\alpha}(-p)\hat{\bar h}_{\rho\beta}(p)+\hat{\bar h}_{\mu\beta}(-p)p^\alpha\hat{\bar h}_{\rho\alpha}(p)\right]\\
		&=i\epsilon^{\mu\nu\rho}p_\nu\left[-p_\mu \tfrac{p^\alpha p^\lambda}{p^2}\hat{\bar h}_{\lambda\alpha}(-p)\hat{\bar h}_{\rho\beta}(p)+\hat{\bar h}_{\mu\beta}(-p)p_\rho \tfrac{p^\alpha p^\lambda}{p^2}\hat{\bar h}_{\lambda\alpha}(p)\right]\\
		&=-i\cancel{\epsilon^{\mu\nu\rho}p_\nu p_\mu} \tfrac{p^\alpha p^\lambda}{p^2}\hat{\bar h}_{\lambda\alpha}(-p)\hat{\bar h}_{\rho\beta}(p)+i\cancel{\epsilon^{\mu\nu\rho}p_\nu p_\rho}\hat{\bar h}_{\mu\beta}(-p) \tfrac{p^\alpha p^\lambda}{p^2}\hat{\bar h}_{\lambda\alpha}(p)=0\ ,
		\end{split}
	\ee
where we used the EoM \eqref{3cond2traceless}.

\section{Summary and conclusions} \label{sec-conclusion}

Fracton models in 3D are widely studied nowadays and find applications in many different contexts, even more than their 4D counterparts. For instance, the duality with elasticity \cite{Pretko:2017kvd} is exclusive of two spatial dimensions. However, the Literature is scattered and lacks a thorough analysis of the theory from a more fundamental point of view. Thus, in this paper, following the QFT approach of \cite{Blasi:2022mbl,Bertolini:2022ijb,Bertolini:2023juh}, we studied the 3D theory of a rank-2 symmetric tensor field with mass dimension 1 which is generated by the covariant fracton symmetry \eqref{fracton symm}. The theory turns out to be a higher rank Chern-Simons theory with a mild dependence on the metric, hence, differently from the standard Chern-Simons, it is not topological. To summarize, the higher rank Chern-Simons theory studied in this paper describes a traceless Hall-like theory of fractons  :
	\bi
	\item{\bf traceless :} even though we started from a traceful rank-2 symmetric tensor $h_{\mu\nu}(x)$, it turns out that the invariant action $S_{inv}$ \eqref{Sinv1traceless} only depends on its traceless component $\bar h_{\mu\nu}(x)$. The traceless nature of our theory is also confirmed by the study of the propagators, for which those involving the trace vanish. This also reflects in the study of the DoF of the theory, to which the trace $h(x)$ does not contributes, with the outcome that the number of the DoF is two.
	\item{\bf Hall-like :} the similarity with the abelian Chern-Simons theory is not only formal. Indeed, as in the 4D case, an electromagnetic analogy emerges from the definition of the traceless invariant field strength $\bar F^{\mu\nu\rho}(x)$ \eqref{defFtraceless}. Thanks to this, we can identify the traceless symmetric rank-2 electric tensor field $\bar E^{ij}(x)$ \eqref{defE1} and the magnetic vector field $B^i(x)$ \eqref{defB1} by looking at the EoM of the theory  as a rank-2 extension of the  Chern-Simons' ones \cite{Dunne:1998qy,Tong:2016kpv}. However, in the standard Chern-Simons theory the full physical content arises only when the theory is coupled to  matter, so that the features of the Hall effect can be recovered. Here the same happens, and a Hall-like effect is observed as well. In fact, from the electric and magnetic -like fields previously identified, a vectorial equation of ``magnetic flux attachment'' \eqref{J=B} is found, together with a Hall-like conductivity tensor $\bar\sigma^{ijkl}$ \eqref{sigma tilde} which ensures the orthogonality relation \eqref{JE=0} between the current $\bar J^{ij}(x)$ \eqref{HallJ} and the electric tensor field $\bar E^{ij}(x)$ \eqref{defE1}. A coefficient similar to this one was also identified in \cite{Pretko:2017xar,Prem:2017kxc}.
	\item{\bf fractons : }the covariant fracton  symmetry \eqref{fracton symm} gives rise to a fractonic theory. The first sign comes from the presence of the higher rank electric and magnetic-like fields \eqref{defE1} and \eqref{defB1}. However, the confirmation is given by the immobility property that defines fractons. Here that property does not come from a Gauss-like constraint (which is absent also in the standard Chern-Simons theory), but it is ensured by the fractonic continuity equation \eqref{cont-fratt} that is recovered from the geometric identity of the EoM \eqref{J=eom'} of theory coupled to matter. Therefore, as in the standard Chern-Simons gauge theory, the physics arises when matter comes into play. This result is not only a confirmation that the theory has fractonic features, but also allows us to identify the dipoles $d^i(x)$ \eqref{dipole def} and the fracton charges $\rho(x)$ \eqref{fracton charge} as  the mobile and immobile objects of the model, respectively. This allows us to claim that the flux attachment relation \eqref{dipole attach} involves the mobile quasiparticles, $i.e.$ the dipoles $d^i(x)$ \eqref{dipole def}, and the higher rank Hall conductivity equation \eqref{HallJ} is related to the dipole current $\bar J^{ij}(x)$ \eqref{J=E}.
	\ei
These results open the doors to many applications and further investigations. The first is to consider the Maxwell-Chern-Simons version of this theory, $i.e.$ the case in which $[h_{\mu\nu}]=1/2$, where the Chern-Simons-like coupling constant should play the role of a mass for the symmetric tensor field $h_{\mu\nu}(x)$, reproducing, for a non-topological theory, the Jackiw-Teitelboim topological mass mechanism. Additionally one could also think of this 3D model as generated by a higher dimensional theta-term, similar to \cite{Bertolini:2023sqa}. Therefore we could expect a Witten-like effect, as in \cite{Pretko:2017xar}, due to a modification of the bulk Gauss constraint related to the presence of a magnetic-like contribution to the charge, as it happens in the standard case. In this context, if we interpret $\rho(x)$ \eqref{fracton charge} as the ``electric'' charge density of the Gauss-like constraint \eqref{constraint}, then the EoM \eqref{defrho} tells us that the theory studied in this paper describes particles carrying both electric and magnetic charges, \textit{i.e.} dyons \cite{Schwinger:1969ib,Witten:1979ey}. Hence, the fractons appearing in our theory exhibit also a dyonic behaviour, which should be investigated in more detail in the future. Furthermore, the flux attachment relation for dipoles \eqref{dipole attach} can be seen as a first step towards a generalization of the particle-vortex duality  \cite{Dunne:1998qy,Jackiw:1990aw} for fractons. In order to do this, one has to introduce the full matter part, thinking about a fracton generalization of the Chern-Simons-Higgs theory, which is the starting point of the full duality. However, in the context of fracton models this is not the only duality that could arise, in fact we already mentioned the one that involves the theory of elasticity \cite{Pretko:2017kvd,Pretko:2019omh}, which is possible only in 3D due to the vectorial nature of the magnetic-like field. It will thus be interesting to understand if the dictionary can be applied also in the 3D model presented in our paper. An insight could come from \cite{Gromov:2017vir} where, in order to study topological elasticity with relations to fractons, a Chern-Simons torsional theory is studied and, under the condition of area preserving diffeomorphisms \cite{Du:2021pbc}, a non covariant fractonic Chern-Simons-like action (whose expression is very close to our covariant one \eqref{Sinv1traceless}) is recovered. If there is indeed a duality in our case, it should involve the generalized Hall response equation \eqref{HallJ} due to the fact that the electric tensor field, according to the dictionary \cite{Pretko:2017kvd}, is directly related to the elastic strain tensor $u^{ij}(x)$, which may thus give an elastic interpretation of the $\bar \sigma^{abcd}$ tensor. Equally interesting is the starting point of \cite{Gromov:2017vir}, $i.e.$ torsional Chern-Simons, which highlights the possibility that fractons are related to torsion. More precisely, in that model the full theory refers to a non-symmetric rank-2 tensor field related to the vielbein \cite{Hughes:2011hv}, and whose transformation contains the fractonic symmetry as a particular case. It could therefore be interesting to investigate this more general model in our QFT framework, in order to understand more deeply the relation between fractons, gravity, torsion and elasticity.

\section*{Acknowledgments}

We thank Alessio Caddeo, Matteo Carrega, Silvia Fasce, Daniele Musso, Giandomenico Palumbo and Davide Rovere for enlightening discussions. This work has been partially supported by the INFN Scientific Initiative GSS: ``Gauge Theory, Strings and Supergravity''. 
\appendix

\section{Level quantization} \label{appQuant}
\subsection{Flux quantization}

The magnetic flux quantization is a crucial point. Indeed, on the solution \eqref{h0isol} of the EoM of $h_{00}(x)$ \eqref{eomF00}, we have that the magnetic field $B^a(x)$ \eqref{defB1} writes as a curl :
	\be\label{Bsol}
	B^a=\epsilon^{0mn}\partial_m\left[h^a_{\ n}-\delta^a_n\left(h^0_{\ 0}+\partial_0\psi\right)\right]\ .
	\ee
Notice that one could for instance make the gauge-fixing choice $h^0_{\ 0}=-\partial_0\psi$, which extends \eqref{h0isol} to the covariant form 
	\be
	h_{\mu0}=\partial_\mu\psi\ ,
	\ee
such that the magnetic field $B^a(x)$ \eqref{Bsol} simplifies to
	\be
	B^a=\epsilon^{0mn}\partial_mh^a_{\ n}\ ,
	\ee
however that is not necessary. Therefore one can use Stokes' theorem for the flux on a surface $\Sigma$ whose boundary is a closed curve $\gamma$
	\be
	\int_\Sigma d^2x B^a=\int_\gamma dx^i\left[h^a_{\ i}-\delta^a_i\left(h^0_{\ 0}+\partial_0\psi\right)\right]\ .
	\ee
The minimum is reached when the integral is done on $\Sigma=S^2$, as done in \cite{Prem:2017kxc,Huang:2023zhp} or \cite{Tong:2016kpv} for standard CS, for which the integral over the boundary $\gamma$ only contributes as a $2\pi$ phase. Thus 
	\be\label{Bquant}
	\int_\Sigma d^2x B^a=2\pi\hat x^a\ ,	
	\ee
where $\hat x^a$ is the unit vector of the direction of the dipole through \eqref{dipole attach}. Notice that this reasoning works if we are considering surfaces with genus zero, and in the continuum, which is our case. However for manifolds of different genus or on the lattice (and their continuum limits) the curve $\gamma$ is not necessarily contractible to give a phase contribution. In these cases singular or discontinuous field configurations must be taken into account, as done for instance in \cite{Seiberg:2020wsg,Seiberg:2020bhn,You:2019bvu,Slagle:2020ugk}.

\subsection{Shift of the action and the large gauge transformation}
In order to discuss the level quantization as done in \cite{Prem:2017kxc,Tong:2016kpv,Huang:2023zhp} the flux of the magnetic field must be taken into account, whose quantization condition, as discussed above, is \eqref{Bquant}. The definition of the fractonic electric and magnetic fields \eqref{defE1} and \eqref{defB1}, and the whole electromagnetic interpretation of section \ref{sec-currents} emerges on the solution of the EoM of $h_{00}(x)$ \eqref{simmF}. This solution in terms of the field strength $\bar F_{\mu\nu\rho}(x)$ \eqref{defFtraceless} can be used interchangeably with the one in terms of the gauge field \eqref{h0isol}. Notice that this coincides with the construction done in \cite{Pretko:2016kxt,Pretko:2016lgv,Prem:2017kxc}, where the scalar field is introduced from the beginning, while here we motivate it from a covariant construction as a solution of an EoM. On the invariant action $S_{inv}$ \eqref{Sinv1} this implies
	\be\label{Sinv(h)}
	S_{inv}|_{h_{0m}=\partial_m\psi}=\int d^3x\epsilon^{\mu\nu\rho}h_\mu^{\ \lambda} \partial_\nu h_{\rho\lambda}=\int d^3x\epsilon^{0mn}\left(2\partial^l\psi\partial_m h_{nl}-h_{ml}\partial_0h_n^{\ l}\right)\ .
	\ee
We can thus consider the thermal partition function in the euclidean periodic time $\tau\sim\tau+\beta$, where $\beta $ is the inverse of the temperature, and define the large gauge transformation defined by the following finite gauge parameter
	\be\label{phi'}
	\phi'(\tau,\vec x)\equiv\frac{2\pi}{\beta}\tau\,\hat x_i x^i\ ,
	\ee
such that under the fracton symmetry \eqref{dfract1} the tensor field $h_{\mu\nu}(x)$ and its generalized field strength $F_{\mu\nu\rho}(x)$ transforms as
	\be\label{large h}
	h_{00}\to h_{00}\quad;\quad h_{0m}\to h_{0m}+\frac{2\pi}{\beta}\hat r_m\quad;\quad h_{mn}\to h_{mn}\quad;\quad F_{\mu\nu\rho}\to F_{\mu\nu\rho}\ .
	\ee
Notice that \eqref{phi'} is not single valued on the periodic time, $i.e.\ \phi'(\tau)\neq\phi'(\tau+\beta)$, however this does not reflect on the transformed fields \eqref{large h}, as happens in the standard case of CS theory \cite{Tong:2016kpv}. Notice also that the transformation \eqref{phi'} is the of the same kind as the one considered in \cite{Prem:2017kxc}, and can be reconduced to the one in \cite{Huang:2023zhp} by taking its gradient $-\partial^a\phi'$. Keeping in mind the definition of the magnetic field \eqref{defB1} and the definition of the traceless field strength $\bar F_{\mu\nu\rho}(x)$ \eqref{defFtraceless}, we can rewrite the invariant action \eqref{Sinv1}, expliciting the coupling constant $k$, as follows :
	\be\label{Sinv(B)}
		\begin{split}
		S_{inv}=&\frac{k}{3}\int d^3x \epsilon^{\mu\nu\rho}h_\mu^{\ \lambda}\bar F_{\lambda\nu\rho}\\
		=&\frac{k}{3}\int d^3x\epsilon^{0mn}\left(h_0^{\ 0}\bar F_{0mn}+h_0^{\ l}\bar F_{lmn}-h_n^{\ 0}\bar F_{00n}-h_m^{\ l}\bar F_{l0n}+h_m^{\ 0}\bar F_{0n0}+h_{m}^{\ l}\bar F_{ln0}\right)\\
		=&\frac{k}{3}\int d^3x\left\{\epsilon^{0mn}\left[h_0^{\ 0}\bar F_{0mn}-h_m^{\ l}\left(\bar F_{0ln}-\bar F_{ln0}\right)\right]+3h_{0l}B^l\right\}\\
		=&k\int d^3x\left[\epsilon^{0mn}\left(h_m^{\ l}\partial_0h_{nl}+h_{0m}\partial_0h_{0n}\right)+2h_{0l}B^l\right]\\
		=&k\int d^3x\left(\epsilon^{0mn}h_m^{\ l}\partial_0h_{nl}+2h_{0l}B^l\right)\ .
		\end{split}
	\ee
where we used the tracelessness and ciclicity properties of $\bar F_{\mu\nu\rho}(x)$ \eqref{TrF=0}, \eqref{ciclicitaF}, the definition of $B^l(x)$ \eqref{defB1} and, in the last line, the solution \eqref{h0isol}. Under the large gauge transformation \eqref{large h} we have that the action transforms as
	\be
	S_{inv}\ \to\ S_{inv}+2k\int d^3x\frac{2\pi}{\beta}\hat x_lB^l\ ,
	\ee
which, evaluated on the sphere $S^2$ using the quantization condition \eqref{Bquant}, finally yields
	\be
	S_{inv}\ \to\ S_{inv}+8\pi^2k\ ,
	\ee
which implies that $k$ must be quantized as 
	\be\label{quant-cond}
	8\pi^2k=2\pi n\quad\Rightarrow\quad k=\frac{1}{4\pi}n\ ,\quad n\in\mathbb{Z}\ ,
	\ee
which is analogous to what happens in ordinary 3D CS theory \cite{Witten:1988hf} with the remarkable difference that the theory we are considering here is not topological.

\subsection{An example on the torus}
We have seen that the invariant action $S_{inv}|_{h_{0m}=\partial_m\psi}$ can be written in two ways, either as
	\be
	S_{inv}|_{h_{0m}=\partial_m\psi}=k\int d^3x\epsilon^{0mn}\left(2\partial^l\psi\partial_m h_{nl}-h_{ml}\partial_0h_n^{\ l}\right)\ ,
	\ee
from \eqref{Sinv(h)}, or
	\be
	S_{inv}|_{h_{0m}=\partial_m\psi}=k\int d^3x\left(-\epsilon^{0mn}h_m^{\ l}\partial_0h_{nl}+2\partial_l\psi B^l\right)\ ,
	\ee
from \eqref{Sinv(B)}. On the large gauge transformation \eqref{large h} the $\partial_0$ contribution has no role in changing the action, so that we can equivalently use
	\be\label{equiv}
	2k\int d^3x\partial^l\psi\epsilon^{0mn}\partial_m h_{nl}\qquad\mbox{or}\qquad2k\int d^3x\partial_l\psi B^l\ .
	\ee
Therefore the flux of the magnetic field is equivalent to the flux of the curl of the gauge field, as in standard electromagnetism. As a consequence of the magnetic-dipole relation \eqref{dipole attach} and the symmetry of the gauge field, we have that
	\be\label{hxx}
	h_{ab}=H\,\hat x_a\hat x_b\ .
	\ee
We can for instance consider a dipole oriented along the $x_1$-axis, $i.e.\ \hat x_1=1,\ \hat x_2=0$, which implies that
	\be
	h_{11}=H\quad;\quad h_{12}=h_{21}=h_{22}=0\ ,
	\ee
as a consequence of \eqref{hxx}. We want to consider an example of quantization on a surface of nonzero genus, such as the torus $T^2$ with periodicity $x_1\sim x_1+l_1$ and $x_2\sim x_2+l_2$, as done in \cite{Seiberg:2020wsg,Seiberg:2020bhn}. The only nontrivial component of the magnetic field is $B^1(x)$, for which we have that the flux on the torus is
	\be\label{torus flux}
	\int_{T^2}d^2xB_1=\int_{T^2}d^2x\partial_2h_{11}
	\ee
as a consequence of the equivalence \eqref{equiv}. We shall see that the magnetic field is quantized if we consider, for instance, the following transition function
	\be
	g(x_1)=\pi\left[(x_1-x_1^*)\theta(x_1-x_1^*)-\frac{x_1}{l_1}(x_1-x_1^*)+\frac{3}{2}\frac{(x_1)^2}{l_1}\right]\ ,
	\ee
which is similar to those used in  \cite{Seiberg:2020wsg,Seiberg:2020bhn}. We thus have
	\be
	h_{11}(x_1,l_2)-h_{11}(x_1,0)=\partial_1\partial_1 g\ ,
	\ee
$i.e.\ h_{11}$ is nonperiodic only around the $x_2$-coordinate. For instance
	\be\label{h11}
	h_{11}=\pi\frac{x_2}{l_2}\left[\delta(x_1-x_1^*)+\frac{1}{l_1}\right]\ .
	\ee
Therefore from \eqref{torus flux} we get
	\be\label{Bflux}
		\begin{split}
		\int_{T^2}d^2xB_1&=\int_{T^2}d^2x\partial_2h_{11}\\
		&=\int_0^{l_1} dx_1\int_0^{l_2} dx_2\left\{\pi\frac{1}{l_2}\left[\delta(x_1-x_1^*)+\frac{1}{l_1}\right]\right\}\\
		&=\pi\int_0^{l_1} dx_1\left[\delta(x_1-x_1^*)+\frac{1}{l_1}\right]\\
		&=2\pi\ .
		\end{split}
	\ee
Therefore in this configuration \eqref{h11}, from \eqref{Bflux} and taking the large gauge transformation \eqref{large h}, the action changes as follows
	\be
	S\ \to\ S+2k\int_0^\beta d\tau\,\frac{2\pi}{\beta}\int_{T^2}d^2x\, B^1=S+8\pi^2k
	\ee
which give the same quantization condition as \eqref{quant-cond}.

\section{Propagator}\label{appProp}

In momentum space the invariant action $S_{inv}$ \eqref{Sinv1} and the gauge fixing term $S_{gf}$ \eqref{Sgf} read
\bea
S_{inv} &=&\!
\int d^3p\; \hat h_{\mu\nu}(p)
\left[
\frac{i}{4}\left(
\epsilon^{\mu\lambda\alpha}\eta^{\nu\beta} + 
\epsilon^{\nu\lambda\alpha}\eta^{\mu\beta} + 
\epsilon^{\mu\lambda\beta}\eta^{\nu\alpha} + 
\epsilon^{\nu\lambda\beta}\eta^{\mu\alpha} 
\right)
p_\lambda
\right]
\hat h_{\alpha\beta}(-p) \label{Sinvp}
\\
S_{gf} &=&\!\!\!
\frac{-1}{2\xi}\int d^3p\; \hat h_{\mu\nu}(p)
\left(
p^\mu p^\nu p^\alpha p^\beta + 
\kappa p^\mu p^\nu p^2 \eta^{\alpha\beta} +
\kappa p^\alpha p^\beta p^2 \eta^{\mu\nu} +
\kappa^2\eta^{\mu\nu}\eta^{\alpha\beta}p^4
\right)
\hat h_{\alpha\beta}(-p) \label{Sgfp}
\ ,
\eea
where $p^4\equiv(p^2)^2=(p^\mu p_\mu)^2$.
The momentum space gauge fixed action $S_{tot}$ \eqref{Stot} is
\be
S_{tot} = -\frac{1}{4}
\int d^3p\; \hat h_{\mu\nu}(p)
\hat G^{\mu\nu,\alpha\beta}(p)
\hat h_{\alpha\beta}(-p) \ ,
\label{Stotp}\ee
where the tensor $\hat G^{\mu\nu,\alpha\beta}(p)$, defined by the sum of \eqref{Sinvp} and \eqref{Sgfp}, has the following symmetries
\be
\hat G^{\mu\nu,\alpha\beta}(p) = \hat G^{\nu\mu,\alpha\beta}(p) = \hat G^{\mu\nu,\beta\alpha}(p)=\hat G^{\alpha\beta,\mu\nu}(-p)\ ,
\label{simmG}\ee
and can be expanded on the following basis of tensors (displaying these same symmetries): 
\bea
A^{(0)}_{\alpha\beta,\rho\sigma}  &=& \frac{1}{2}(\eta_{\alpha\rho}\eta_{\beta\sigma}+\eta_{\alpha\sigma}\eta_{\beta\rho}) 
\label{A0}\\
A^{(1)}_{\alpha\beta,\rho\sigma} &=& 
\eta_{\alpha\rho}p_\beta p_\sigma +  
\eta_{\alpha\sigma}p_\beta p_\rho +
\eta_{\beta\rho}p_\alpha p_\sigma +
\eta_{\beta\sigma}p_\alpha p_\rho
\label{A1}\\
A^{(2)}_{\alpha\beta,\rho\sigma} &=& 
\eta_{\alpha\beta} p_\rho p_\sigma +
\eta_{\rho\sigma} p_\alpha p_\beta
\label{A2} \\
A^{(3)}_{\alpha\beta,\rho\sigma} &=& 
\eta_{\alpha\beta}\eta_{\rho\sigma}
\label{A3}\\
A^{(4)}_{\alpha\beta,\rho\sigma} &=& 
p_\alpha p_\beta p_\rho p_\sigma
\label{A4}\\
A^{(5)}_{\alpha\beta,\rho\sigma} &=& 
ip^\lambda (
\epsilon_{\alpha\lambda\rho} \eta_{\sigma\beta} +
\epsilon_{\beta\lambda\rho} \eta_{\sigma\alpha} +
\epsilon_{\alpha\lambda\sigma} \eta_{\rho\beta} +
\epsilon_{\beta\lambda\sigma} \eta_{\rho\alpha}
)
\label{A5}\\
A^{(6)}_{\alpha\beta,\rho\sigma} &=&
 ip^\lambda (
\epsilon_{\alpha\lambda\rho} p_\sigma p_\beta +
\epsilon_{\alpha\lambda\sigma} p_\rho p_\beta +
\epsilon_{\beta\lambda\rho} p_\sigma p_\alpha +
\epsilon_{\beta\lambda\sigma} p_\rho p_\alpha
)\ 
\label{A6}
\eea
as follows
\be
\hat G^{\mu\nu,\alpha\beta}(p) = 
\left[
-A^{(5)} +
\frac{2}{\xi} \left(
\kappa p^2 A^{(2)} + \kappa^2p^4 A^{(3)} + A^{(4)}
\right)   \right]^{\mu\nu,\alpha\beta}
\ .
\label{Gexpansion}\ee
The momentum space propagator in a generic $\xi$-gauge 
\be
\hat\Delta^{(\xi)} _{\alpha\beta,\rho\sigma}(p)
\equiv
\langle\hat h_{\alpha\beta}(p)\hat h_{\rho\sigma}(-p)
\rangle^{(\xi)} 
\label{Delta}\ee
is defined as
\be
\hat G \hat \Delta^{(\xi)} ={\cal A}^{(0)} \ .
\label{defDelta}\ee
In the basis $\{A^{(i)}\}$ the propagator $\hat\Delta^{(\xi)}$ reads
\be
\hat \Delta^{(\xi)} = \sum_{i=0}^6 c_i A^{(i)}\ ,
\label{Deltabase}\ee
where the $p$-dependent coefficients $c_i(p)$ are determined by \eqref{defDelta}. In order to solve the equation \eqref{defDelta} the following tensorial relations are useful \cite{Maggiore:2018bxr,Bertolini:2020hgr,Bertolini:2022sao}
\begin{align}
A^{(0)}A^{(0)} &= {\cal A}^{(0)}\label{A0A0}\\
A^{(0)}A^{(1)} &= {\cal A}^{(1)}
\\
A^{(0)}A^{(2)} &= {\cal A}^{(2)} + {\cal B}^{(2)} \\
A^{(0)}A^{(3)} &= {\cal A}^{(3)} \\
A^{(0)}A^{(4)} &= {\cal A}^{(4)} \\
A^{(0)}A^{(5)} &= {\cal A}^{(5)}  \\
A^{(0)}A^{(6)} &= {\cal A}^{(6)} \\
\intertext{}
A^{(1)}A^{(0)} &= {\cal A}^{(1)}
\\
A^{(1)}A^{(1)} &= 2p^2{\cal A}^{(1)}
+8{\cal A}^{(4)} \\
A^{(1)}A^{(2)} &= 4{\cal A}^{(4)} + 4p^2{\cal B}^{(2)} \\
A^{(1)}A^{(3)} &= 4{\cal B}^{(2)} \\
A^{(1)}A^{(4)} &= 4p^2{\cal A}^{(4)} \\
A^{(1)}A^{(5)} &= 2{\cal A}^{(6)}  \\
A^{(1)}A^{(6)} &= 2p^2{\cal A}^{(6)} \\
\intertext{}
A^{(2)}A^{(0)} &= {\cal A}^{(2)} + {\cal B}^{(2)} \\
A^{(2)}A^{(1)} &= 4p^2{\cal A}^{(2)} + 4{\cal A}^{(4)} \\
A^{(2)}A^{(2)} &= p^2({\cal A}^{(2)} + {\cal B}^{(2)}+p^2{\cal A}^{(3)}) + 3 {\cal A}^{(4)} \\
A^{(2)}A^{(3)} &= 3{\cal B}^{(2)} + p^2 {\cal A}^{(3)} \\
A^{(2)}A^{(4)} &=  p^4{\cal A}^{(2)} + p^2{\cal A}^{(4)}\\
A^{(2)}A^{(5)} &= A^{(2)}A^{(6)} = 0\\
\intertext{}
A^{(3)}A^{(0)} &= {\cal A}^{(3)}\\
A^{(3)}A^{(1)} &= 4{\cal A}^{(2)} \\
A^{(3)}A^{(2)} &= 3{\cal A}^{(2)} + p^2{\cal A}^{(3)} \\
A^{(3)}A^{(3)} &= 3{\cal A}^{(3)} \\
A^{(3)}A^{(4)} &= p^2{\cal A}^{(2)} \\
A^{(3)}A^{(5)} &= A^{(3)}A^{(6)} = 0\\
\intertext{}
A^{(4)}A^{(0)} &= {\cal A}^{(4)}\\
A^{(4)}A^{(1)} &= 4p^2{\cal A}^{(4)}  \\
A^{(4)}A^{(2)} &= p^4{\cal B}^{(2)} + p^2{\cal A}^{(4)} \\
A^{(4)}A^{(3)} &= p^2{\cal B}^{(2)} \\
A^{(4)}A^{(4)} &= p^4{\cal A}^{(4)} \\
A^{(4)}A^{(5)} &= A^{(4)}A^{(6)} = 0 \\
\intertext{}
A^{(5)}A^{(0)} &= {\cal A}^{(5)}\\
A^{(5)}A^{(1)} &= 2{\cal A}^{(6)}  \\
A^{(5)}A^{(2)} &= A^{(5)}A^{(3)} = A^{(5)}A^{(4)} = 0 \\
A^{(5)}A^{(5)} &= -16p^2{\cal A}^{(0)} + 6{\cal A}^{(1)}
- 8{\cal A}^{(2)}- 8{\cal B}^{(2)}
+8p^2{\cal A}^{(3)}   \\
A^{(5)}A^{(6)} &= 8{\cal A}^{(4)} - 2p^2{\cal A}^{(1)}
\\
\intertext{}
A^{(6)}A^{(0)} &= {\cal A}^{(6)}\\
A^{(6)}A^{(1)} &= 2p^2{\cal A}^{(6)} \\
A^{(6)}A^{(2)} &= A^{(6)}A^{(3)} = A^{(6)}A^{(4)} =0\\
A^{(6)}A^{(5)} &= -2p^2{\cal A}^{(1)}
+8{\cal A}^{(4)}  \\
A^{(6)}A^{(6)} &=  8p^2{\cal A}^{(4)}-2p^2{\cal A}^{(1)}\label{A6A6}\ ,
\end{align}
where the contraction of indices, which has been omitted, is as follows
\be
(XY={\cal Z}) \equiv (X^{\mu\nu,\alpha\beta}Y_{\alpha\beta,\rho\sigma}={\cal Z}^{\mu\nu}_{\rho\sigma})\ ,
\label{}\ee
and the operators appearing at the right hand side of \eqref{A0A0}-\eqref{A6A6} form a basis on the space of the tensors 
\be
{\cal Z}^{\mu\nu}_{\rho\sigma} = {\cal Z}^{\nu\mu}_{\rho\sigma} = {\cal Z}^{\mu\nu}_{\sigma\rho}\ ,
\label{simmZ}\ee
which is larger than the one concerning the more symmetric tensors \eqref{simmG}:
\bea
{\cal A}^{(0)\mu\nu}_{\quad\rho\sigma} &=& \frac{1}{2}(\delta^\mu_\rho\delta^\nu_\sigma + \delta^\mu_\sigma\delta^\nu_\rho)  \\
{\cal A}^{(1)\mu\nu}_{\quad\rho\sigma} &=& \delta^\mu_\rho p^\nu p_\sigma + \delta^\mu_\sigma p^\nu p_\rho + \delta^\nu_\rho p^\mu p_\sigma + \delta^\nu_\sigma p^\mu p_\rho   \\
{\cal A}^{(2)\mu\nu}_{\quad\rho\sigma} &=& \eta^{\mu\nu} p_\rho p_\sigma  \\
{\cal B}^{(2)\mu\nu}_{\quad\rho\sigma} &=& \eta_{\rho\sigma} p^\mu p^\nu  \\
{\cal A}^{(3)\mu\nu}_{\quad\rho\sigma} &=&  \eta^{\mu\nu} \eta_{\rho\sigma}\\
{\cal A}^{(4)\mu\nu}_{\quad \rho\sigma} &=& p^\mu p^\nu p_\rho p_\sigma\\
{\cal A}^{(5)\mu\nu}_{\quad\rho\sigma} &=& ip^\lambda
(
\epsilon_{\alpha\lambda\rho}\eta^{\mu\alpha}\delta^\nu_\sigma +
\epsilon_{\alpha\lambda\sigma}\eta^{\mu\alpha}\delta^\nu_\rho +
\epsilon_{\alpha\lambda\rho}\eta^{\nu\alpha}\delta^\mu_\sigma +
\epsilon_{\alpha\lambda\sigma}\eta^{\nu\alpha}\delta^\mu_\rho
)\\
{\cal A}^{(6)\mu\nu}_{\quad\rho\sigma} &=& 
ip^\lambda 
(
\eta^{\mu\alpha}\epsilon_{\alpha\lambda\rho}p_\sigma p^\nu +
\eta^{\mu\alpha}\epsilon_{\alpha\lambda\sigma}p_\rho p^\nu +
\eta^{\nu\alpha}\epsilon_{\alpha\lambda\rho}p_\sigma p^\mu +
\eta^{\nu\alpha}\epsilon_{\alpha\lambda\sigma}p_\rho p^\mu
) 
\eea
The usual long but straightforward calculations lead then to
\begin{align}
\hat G \hat \Delta^{(\xi)} = &
16c_5p^2 {\cal A}^{(0)} +(-6c_5+2p^2c_6){\cal A}^{(1)} \nonumber\\
+& \left\{\tfrac{2\kappa p^4}{\xi} 
\left[\tfrac{c_0}{p^2} + 4(\kappa+1)p^2 c_1+(3\kappa+1) c_2 + (\kappa+1) p^2c_4\right]+8c_5\right\} {\cal A}^{(2)}  \nonumber\\
+& 
\left\{ \tfrac{2p^2}{\xi} 
\left[\kappa c_0+(\kappa+1) p^2 c_2 + (3\kappa+1)c_3\right]+8c_5\right\}{\cal B}^{(2)}\nonumber\\ +&
\left\{ \tfrac{2\kappa p^2}{\xi} 
\left[\kappa c_0 +(\kappa+1) p^2 c_2 + (3\kappa+1)c_3\right]-8c_5\right\}p^2{\cal A}^{(3)} 
\nonumber\\
 +&
 \left\{ \tfrac{2p^2}{\xi} 
\left[\tfrac{c_0}{p^2} + 4(\kappa+1) c_1+(3\kappa+1)c_2 + (\kappa+1)p^2c_4\right]-8c_6\right\}{\cal A}^{(4)} -c_0 {\cal A}^{(5)}-2c_1 {\cal A}^{(6)} \nonumber\\
= & \ {\cal A}^{(0)}\ .\label{coeffeq}
\end{align}
The solution of the above equation gives the coefficients $c_i(p)$ appearing in the propagator expansion~\eqref{Deltabase}
\be
\xi=c_0 = c_1 = 0 \ ;\ 
c_2 = \mbox{free} \ ;\ 
c_3 =- c_2\frac{1+\kappa}{1+3\kappa}p^2\ ;\ 
c_4 =-c_2\frac{1+3\kappa}{(1+\kappa)p^2}\ ;\ 
c_5 = \frac{1}{16p^2}\ ;\ 
c_6 = \frac{3}{16p^4}   \ .
\label{}\ee
The propagator displays poles at $\kappa=\{-1,-\frac{1}{3}\}$, and $\kappa$ is coupled to the trace $h(x)$ in the gauge fixing term \eqref{Sgf}, on which, on the other hand, the invariant action $S_{inv}$ \eqref{Sinv1traceless} does not depend. We notice  that $c_2(p)$ is a free parameter, which means that it does not have a role in inverting the matrix \eqref{Gexpansion}, hence we can set it to zero. This also makes the singular coefficients vanish. Additionally, as one might hope and expect, the Landau gauge is a mandatory choice. Indeed, due to the fact that the gauge parameter is massive $([\xi]=3)$, its presence would lead to infrared divergences in the correlation functions. Therefore, the only surviving coefficients are $c_5(p)$ and $c_6(p)$, and the propagator is
\be
\begin{split}
&\langle
\hat h_{\alpha\beta}(p)\hat h_{\rho\sigma}(-p)
\rangle
=\hat\Delta_{\alpha\beta,\rho\sigma}(p)=  \\
&\frac{ip^\lambda}{16p^2}\left[
\epsilon_{\alpha\lambda\rho}
\left(
\eta_{\beta\sigma} + 3 \frac{p_\beta p_\sigma}{p^2}
\right)
+
\epsilon_{\alpha\lambda\sigma}
\left(
\eta_{\beta\rho} + 3 \frac{p_\beta p_\rho}{p^2}
\right)
+
\epsilon_{\beta\lambda\rho}
\left(
\eta_{\alpha\sigma} + 3 \frac{p_\alpha p_\sigma}{p^2}
\right)
+
\epsilon_{\beta\lambda\sigma}
\left(
\eta_{\alpha\rho} + 3 \frac{p_\alpha p_\rho}{p^2}
\right)
\right].
\end{split}
\label{proplandauapp}\ee

\newpage



\begin{thebibliography}{15}

\bibitem{Anderson:1972pca}
P.~W.~Anderson,
Science \textbf{177} (1972) no.4047, 393-396
doi:10.1126/science.177.4047.393.

\bibitem{Sachdev:2012dq}
S.~Sachdev,
``The quantum phases of matter,''
[arXiv:1203.4565 [hep-th]].

\bibitem{Fradkin:2013sab}
E.~H.~Fradkin,
``Field Theories of Condensed Matter Physics,''
Front. Phys. \textbf{82} (2013), 1-852
Cambridge Univ. Press, 2013,
ISBN 978-0-521-76444-5, 978-1-107-30214-3.

\bibitem{Witten:2015aoa}
E.~Witten,
Riv. Nuovo Cim. \textbf{39} (2016) no.7, 313-370
doi:10.1393/ncr/i2016-10125-3.

\bibitem{Vijay:2015mka}
S.~Vijay, J.~Haah and L.~Fu,
Phys. Rev. B \textbf{92} (2015) no.23, 235136
doi:10.1103/PhysRevB.92.235136.

\bibitem{Vijay:2016phm}
S.~Vijay, J.~Haah and L.~Fu,
Phys. Rev. B \textbf{94} (2016) no.23, 235157
doi:10.1103/PhysRevB.94.235157.


\bibitem{Nandkishore:2018sel}
R.~M.~Nandkishore and M.~Hermele,
Ann. Rev. Condensed Matter Phys. \textbf{10}, 295-313 (2019)
doi:10.1146/annurev-conmatphys-031218-013604.



\bibitem{Pretko:2020cko}
M.~Pretko, X.~Chen and Y.~You,
Int. J. Mod. Phys. A \textbf{35} (2020) no.06, 2030003
doi:10.1142/S0217751X20300033.

\bibitem{Caddeo:2022ibe}
A.~Caddeo, C.~Hoyos and D.~Musso,
Phys. Rev. D \textbf{106} (2022) no.11, L111903
doi:10.1103/PhysRevD.106.L111903.


\bibitem{rasmussen}
A.~Rasmussen, Y.-Z.~You and C.~Xu,
doi:10.48550/arXiv.1601.08235
[arXiv:1601.08235v1 [cond-mat.str-el]].

\bibitem{Pretko:2018jbi}
M.~Pretko,
Phys. Rev. B \textbf{98} (2018) no.11, 115134
doi:10.1103/PhysRevB.98.115134.

\bibitem{Chamon:2004lew}
C.~Chamon,
Phys. Rev. Lett. \textbf{94} (2005) no.4, 040402
doi:10.1103/physrevlett.94.040402.

\bibitem{Haah:2011drr}
J.~Haah,
Phys. Rev. A \textbf{83} (2011) no.4, 042330
doi:10.1103/physreva.83.042330.

\bibitem{Bravyi:quantum}
S.~Bravyi, J.~Haah,
Phys. Rev. Lett \textbf{111}, no.20, 200501 (2013)
doi.org/10.1103/PhysRevLett.111.200501.

\bibitem{Yoshida:2013sqa}
B.~Yoshida,
Phys. Rev. B \textbf{88} (2013) no.12, 125122
doi:10.1103/PhysRevB.88.125122.

\bibitem{Pretko:2016kxt}
M.~Pretko,
Phys. Rev. B \textbf{95}, no.11, 115139 (2017)
doi:10.1103/PhysRevB.95.115139.

\bibitem{Pretko:2017kvd}
M.~Pretko and L.~Radzihovsky,
Phys. Rev. Lett. \textbf{120} (2018) no.19, 195301
doi:10.1103/PhysRevLett.120.195301.


\bibitem{Pretko:2019omh}
M.~Pretko, Z.~Zhai and L.~Radzihovsky,
Phys. Rev. B \textbf{100} (2019) no.13, 134113
doi:10.1103/PhysRevB.100.134113.

\bibitem{Lucas}
A.~Gromov, A.~Lucas and R.~M.~Nandkishore,
Phys. Rev. Res. \textbf{2}, no.3, 033124 (2020)
doi:10.1103/PhysRevResearch.2.033124.

\bibitem{Ye}
J.~K.~Yuan, S.~A.~Chen and P.~Ye,
Phys. Rev. Res. \textbf{2}, no.2, 023267 (2020)
doi:10.1103/PhysRevResearch.2.023267.

\bibitem{Wang}
J.~Wang and S.~T.~Yau,
Phys. Rev. Res. \textbf{2}, no.4, 043219 (2020)
doi:10.1103/PhysRevResearch.2.043219.

\bibitem{Surowska2}
K.~T.~Grosvenor, C.~Hoyos, F.~Pe\~na-Ben\'\i{}tez and P.~Sur\'owka,
Phys. Rev. Res. \textbf{3}, no.4, 043186 (2021)
doi:10.1103/PhysRevResearch.3.043186.

\bibitem{Surowska3}
A.~G\l{}\'odkowski, F.~Pe\~na-Ben\'\i{}tez and P.~Sur\'owka,
Phys. Rev. E \textbf{107}, no.3, 034142 (2023)
doi:10.1103/PhysRevE.107.034142.

\bibitem{Yan:2018nco}
H.~Yan,
Phys. Rev. B \textbf{99} (2019) no.15, 155126
doi:10.1103/PhysRevB.99.155126.

\bibitem{Gu:2006vw}
Z.~C.~Gu and X.~G.~Wen,
[arXiv:gr-qc/0606100 [gr-qc]].

\bibitem{Xu:2006}
C.~Xu,
Phys. Rev. B \textbf{74} (2006) no.22, 224433
doi:10.1103/PhysRevB.74.224433.

\bibitem{Gu:2009jh}
Z.~C.~Gu and X.~G.~Wen,
Nucl. Phys. B \textbf{863} (2012), 90-129
doi:10.1016/j.nuclphysb.2012.05.010.

\bibitem{Xu:2010eg}
C.~Xu and P.~Horava,
Phys. Rev. D \textbf{81} (2010), 104033
doi:10.1103/PhysRevD.81.104033.

\bibitem{Shirley:2017suz}
W.~Shirley, K.~Slagle, Z.~Wang and X.~Chen,
Phys. Rev. X \textbf{8} (2018) no.3, 031051
doi:10.1103/PhysRevX.8.031051.

\bibitem{Pretko:2017fbf}
M.~Pretko,
Phys. Rev. D \textbf{96}, no.2, 024051 (2017)
doi:10.1103/PhysRevD.96.024051.

\bibitem{Jain:2021ibh}
A.~Jain and K.~Jensen,
SciPost Phys. \textbf{12} (2022) no.4, 142
doi:10.21468/SciPostPhys.12.4.142.

\bibitem{Tsaloukidis:2023bvz}
L.~Tsaloukidis, J.~J.~Fern\'andez-Melgarejo, J.~Molina-Vilaplana and P.~Sur\'owka,
[arXiv:2304.12242 [hep-th]].

\bibitem{Pretko:2016lgv}
M.~Pretko,
Phys. Rev. B \textbf{96} (2017) no.3, 035119
doi:10.1103/PhysRevB.96.035119.

\bibitem{Ma:2018nhd}
H.~Ma, M.~Hermele and X.~Chen,
Phys. Rev. B \textbf{98} (2018) no.3, 035111
doi:10.1103/PhysRevB.98.035111.

\bibitem{Bulmash:2018lid}
D.~Bulmash and M.~Barkeshli,
Phys. Rev. B \textbf{97} (2018) no.23, 235112
doi:10.1103/PhysRevB.97.235112.

\bibitem{Seiberg:2020wsg}
N.~Seiberg and S.~H.~Shao,
SciPost Phys. \textbf{9} (2020) no.4, 046
doi:10.21468/SciPostPhys.9.4.046.

\bibitem{Seiberg:2020bhn}
N.~Seiberg and S.~H.~Shao,
SciPost Phys. \textbf{10} (2021) no.2, 027
doi:10.21468/SciPostPhys.10.2.027.

\bibitem{Seiberg:2020cxy}
N.~Seiberg and S.~H.~Shao,
SciPost Phys. \textbf{10} (2021) no.1, 003
doi:10.21468/SciPostPhys.10.1.003.

\bibitem{Blasi:2022mbl}
A.~Blasi and N.~Maggiore,
Phys. Lett. B C \textbf{833}, 137304 (2022)
doi:10.1016/j.physletb.2022.137304.

\bibitem{Bertolini:2022ijb}
E.~Bertolini and N.~Maggiore,
Phys. Rev. D \textbf{106}, no.12, 125008 (2022)
doi:10.1103/PhysRevD.106.125008.

\bibitem{Bertolini:2023juh}
E.~Bertolini, A.~Blasi, A.~Damonte and N.~Maggiore,
Symmetry \textbf{15} (2023) no.4, 945
doi:10.3390/sym15040945.

\bibitem{Afxonidis:2023pdq}
E.~Afxonidis, A.~Caddeo, C.~Hoyos and D.~Musso,
[arXiv:2311.01818 [hep-th]].

\bibitem{Dalmazi:2020xou}
D.~Dalmazi and R.~R.~L.~d.~Santos,
Eur. Phys. J. C \textbf{81} (2021) no.6, 547
doi:10.1140/epjc/s10052-021-09297-0.

\bibitem{Hinterbichler:2011tt}
K.~Hinterbichler,
Rev. Mod. Phys. \textbf{84} (2012), 671-710
doi:10.1103/RevModPhys.84.671.

\bibitem{Bertolini:2023wie}
E.~Bertolini and N.~Maggiore,
Phys. Rev. D \textbf{108} (2023) no.10, 105012
doi:10.1103/PhysRevD.108.105012.

\bibitem{Amoretti:2014iza}
A.~Amoretti, A.~Braggio, G.~Caruso, N.~Maggiore and N.~Magnoli,
Phys. Rev. D \textbf{90} (2014) no.12, 125006
doi:10.1103/PhysRevD.90.125006.

\bibitem{You:2019bvu}
Y.~You, F.~J.~Burnell and T.~L.~Hughes,
Phys. Rev. B \textbf{103} (2021) no.24, 245128
doi:10.1103/PhysRevB.103.245128.

\bibitem{Bertolini:2023sqa}
E.~Bertolini, N.~Maggiore and G.~Palumbo,
Phys. Rev. D \textbf{108}, no.2, 025009 (2023)
doi:10.1103/PhysRevD.108.025009..

\bibitem{Pretko:2017xar}
M.~Pretko,
Phys. Rev. B \textbf{96}, no.12, 125151 (2017)
doi:10.1103/PhysRevB.96.125151.

\bibitem{Prem:2017kxc}
A.~Prem, M.~Pretko and R.~Nandkishore,
Phys. Rev. B \textbf{97} (2018) no.8, 085116
doi:10.1103/PhysRevB.97.085116.

\bibitem{Delfino:2022ndx}
G.~Delfino, W.~B.~Fontana, P.~R.~S.~Gomes and C.~Chamon,
SciPost Phys. \textbf{14} (2023) no.1, 002
doi:10.21468/SciPostPhys.14.1.002.

\bibitem{You:2019ciz}
Y.~You, T.~Devakul, S.~L.~Sondhi and F.~J.~Burnell,
Phys. Rev. Res. \textbf{2} (2020) no.2, 023249
doi:10.1103/PhysRevResearch.2.023249.

\bibitem{Delfino:2023anb}
G.~Delfino and Y.~You,
Phys. Rev. B \textbf{109} (2024) no.20, 205146
doi:10.1103/PhysRevB.109.205146.

\bibitem{Fliss:2021ekk}
J.~R.~Fliss,
SciPost Phys. \textbf{11} (2021) no.3, 052
doi:10.21468/SciPostPhys.11.3.052.

\bibitem{Cappelli:2015ocj}
A.~Cappelli and E.~Randellini,
JHEP \textbf{03} (2016), 105
doi:10.1007/JHEP03(2016)105.

\bibitem{Dunne:1998qy}
G.~V.~Dunne,
``Aspects of Chern-Simons theory,''
Contribution to: Les Houches Summer School in Theoretical Physics, Session 69: Topological Aspects of Low-dimensional Systems, 
7-31 July 1998. Les Houches, France
[arXiv:hep-th/9902115 [hep-th]].

\bibitem{Tong:2016kpv}
D.~Tong,
``Lectures on the Quantum Hall Effect,''
http://www.damtp.cam.ac.uk/user/tong/qhe.html
[arXiv:1606.06687 [hep-th]].

\bibitem{Bergshoeff:2009tb}
E.~A.~Bergshoeff, O.~Hohm and P.~K.~Townsend,
Annals Phys. \textbf{325} (2010), 1118-1134
doi:10.1016/j.aop.2009.12.010.

\bibitem{Wu:1988py}
Y.~S.~Wu and A.~Zee,
Phys. Lett. B \textbf{207} (1988), 39-43
doi:10.1016/0370-2693(88)90882-9.

\bibitem{Gromov:2017vir}
A.~Gromov,
Phys. Rev. Lett. \textbf{122} (2019) no.7, 076403
doi:10.1103/PhysRevLett.122.076403.

\bibitem{Deser:1981wh}
S.~Deser, R.~Jackiw and S.~Templeton,
Annals Phys. \textbf{140}, 372-411 (1982)
[erratum: Annals Phys. \textbf{185}, 406 (1988)]
doi:10.1016/0003-4916(82)90164-6.

\bibitem{Huang:2023zhp}
X.~Huang,
SciPost Phys. \textbf{15}, no.4, 153 (2023)
doi:10.21468/SciPostPhys.15.4.153.

\bibitem{Jackiw:1990aw}
R.~Jackiw and E.~J.~Weinberg,
Phys. Rev. Lett. \textbf{64} (1990), 2234
doi:10.1103/PhysRevLett.64.2234.

\bibitem{Alvarez-Gaume:1989ldl}
L.~Alvarez-Gaume, J.~M.~F.~Labastida and A.~V.~Ramallo,
Nucl. Phys. B \textbf{334}, 103-124 (1990)
doi:10.1016/0550-3213(90)90658-Z.

\bibitem{Blasi:1989mw}
A.~Blasi and R.~Collina,
Nucl. Phys. B \textbf{345}, 472-492 (1990)
doi:10.1016/0550-3213(90)90397-V.

\bibitem{Delduc:1990je}
F.~Delduc, C.~Lucchesi, O.~Piguet and S.~P.~Sorella,
Nucl. Phys. B \textbf{346}, 313-328 (1990)
doi:10.1016/0550-3213(90)90283-J.

\bibitem{Blasi:2015lrg}
A.~Blasi and N.~Maggiore,
Class. Quant. Grav. \textbf{34} (2017) no.1, 015005
doi:10.1088/1361-6382/34/1/015005.

\bibitem{Blasi:2017pkk}
A.~Blasi and N.~Maggiore,
Eur. Phys. J. C \textbf{77} (2017) no.9, 614
doi:10.1140/epjc/s10052-017-5205-y.

\bibitem{Gambuti:2020onb}
G.~Gambuti and N.~Maggiore,
Phys. Lett. B \textbf{807} (2020), 135530
doi:10.1016/j.physletb.2020.135530.

\bibitem{Gambuti:2021meo}
G.~Gambuti and N.~Maggiore,
Eur. Phys. J. C \textbf{81} (2021) no.2, 171
doi:10.1140/epjc/s10052-021-08962-8.

\bibitem{Bertolini:2021iku}
E.~Bertolini, G.~Gambuti and N.~Maggiore,
Phys. Rev. D \textbf{104} (2021) no.10, 105011
doi:10.1103/PhysRevD.104.105011.

\bibitem{Maggiore:2017vjf}
N.~Maggiore,
Int. J. Mod. Phys. A \textbf{33} (2018) no.02, 1850013
doi:10.1142/S0217751X18500136.

\bibitem{Birmingham:1991ty}
D.~Birmingham, M.~Blau, M.~Rakowski and G.~Thompson,
Phys. Rept. \textbf{209} (1991), 129-340
doi:10.1016/0370-1573(91)90117-5.

\bibitem{Schwinger:1969ib}
J.~S.~Schwinger,
Science \textbf{165} (1969), 757-761
doi:10.1126/science.165.3895.757.

\bibitem{Witten:1979ey}
E.~Witten,
Phys. Lett. B \textbf{86} (1979), 283-287
doi:10.1016/0370-2693(79)90838-4.

\bibitem{Du:2021pbc}
Y.~H.~Du, U.~Mehta, D.~X.~Nguyen and D.~T.~Son,
SciPost Phys. \textbf{12} (2022) no.2, 050
doi:10.21468/SciPostPhys.12.2.050.

\bibitem{Hughes:2011hv}
T.~L.~Hughes, R.~G.~Leigh and E.~Fradkin,
Phys. Rev. Lett. \textbf{107} (2011), 075502
doi:10.1103/PhysRevLett.107.075502.

\bibitem{Slagle:2020ugk}
K.~Slagle,
Phys. Rev. Lett. \textbf{126} (2021) no.10, 101603
doi:10.1103/PhysRevLett.126.101603.

\bibitem{Witten:1988hf}
E.~Witten,
Commun. Math. Phys. \textbf{121} (1989), 351-399
doi:10.1007/BF01217730

\bibitem{Maggiore:2018bxr}
N.~Maggiore,
Eur. Phys. J. Plus \textbf{133} (2018) no.7, 281
doi:10.1140/epjp/i2018-12130-y.

\bibitem{Bertolini:2020hgr}
E.~Bertolini and N.~Maggiore,
Symmetry \textbf{12} (2020) no.7, 1134
doi:10.3390/sym12071134.

\bibitem{Bertolini:2022sao}
E.~Bertolini, F.~Fecit and N.~Maggiore,
Symmetry \textbf{14} (2022) no.4, 675
doi:10.3390/sym14040675.







\end{thebibliography}
\end{document}